\documentclass[aps,prd,twocolumn,superscriptaddress,raggedbottom,showpacs,preprintnumbers,floatfix]{revtex4}
\usepackage{bm}
\usepackage{graphicx}

\newcommand{\ex}{\widehat{\bm{x}}}
\newcommand{\ey}{\widehat{\bm{y}}}
\newcommand{\ez}{\widehat{\bm{z}}}
\newcommand{\xvec}{\bm{x}}
\newcommand{\bvec}{\bm{b}}

\newcommand{\pvec}{\bm{p}}

\begin{document}
\preprint{TCDMATH 13--02}
\title{Extended hadron and two-hadron operators of definite momentum for \\
       spectrum calculations in lattice QCD}
\author{C.~Morningstar}
\affiliation{Department of Physics, 
             Carnegie Mellon University, 
             Pittsburgh, Pennsylvania 15213, USA}
\author{J.~Bulava}
\affiliation{School of Mathematics, 
             Trinity College, Dublin 2, Ireland}
\affiliation{CERN, Physics Department,
             CH-1211 Geneva 23, Switzerland}
\author{B.~Fahy}
\affiliation{Department of Physics, 
             Carnegie Mellon University, 
             Pittsburgh, Pennsylvania 15213, USA}
\author{J.~Foley}
\affiliation{Department of Physics and Astronomy, 
             University of Utah, 
             Salt Lake City, Utah 84112, USA}
\author{Y.C.~Jhang}
\affiliation{Department of Physics, 
             Carnegie Mellon University, 
             Pittsburgh, Pennsylvania 15213, USA}
\author{K.J.~Juge}
\affiliation{Department of Physics, 
             University of the Pacific, 
             Stockton, California 95211, USA}
\author{D.~Lenkner}
\affiliation{Department of Physics, 
             Carnegie Mellon University, 
             Pittsburgh, Pennsylvania 15213, USA}
\author{C.H.~Wong}
\affiliation{Department of Physics, 
             University of California at San Diego,
             La Jolla, California 92093, USA} 

\date{June 21, 2013}

\begin{abstract}
Multi-hadron operators are crucial for reliably extracting the masses of 
excited states lying above multi-hadron thresholds in lattice QCD Monte Carlo 
calculations.  The construction of multi-hadron operators with significant
coupling to the lowest-lying multi-hadron states of interest involves combining 
single hadron operators of various momenta.  The design and implementation of 
large sets of spatially-extended single-hadron operators of definite momentum
and their combinations into two-hadron operators are described.  The single
hadron operators are all assemblages of gauge-covariantly-displaced, smeared
quark fields.  Group-theoretical projections onto the irreducible representations 
of the symmetry group of a cubic spatial lattice are used in all isospin channels.
Tests of these operators on $24^3\times 128$ and $32^3\times 256$
anisotropic lattices using a stochastic method of treating the
low-lying modes of quark propagation which exploits Laplacian Heaviside
quark-field smearing are presented.  The method provides reliable estimates
of all needed correlations, even those that are particularly difficult to 
compute, such as $\eta\eta\rightarrow\eta\eta$ in the scalar channel, which
involves the subtraction of a large vacuum expectation value. 
A new glueball operator is introduced, and the evaluation of the mixing
of this glueball operator with a quark-antiquark operator, $\pi\pi$,
and $\eta\eta$ operators is shown to be feasible.
\end{abstract}
\maketitle

\section{Introduction}
Markov-chain Monte Carlo estimates of quantum chromodynamics (QCD) path 
integrals defined on a space-time lattice are a promising means of 
calculating the mass spectrum of excited-state hadron resonances.
Because of the way in which stationary-state energies are extracted from 
the temporal correlations of suitable quantum field operators in such 
calculations, the energy of a particular state of interest can only be 
determined after contributions from all lower-lying and nearby states in the
same symmetry channel are carefully considered. Multi-hadron states populate 
the spectrum below most of the resonances of interest.  To reliably determine
the energies of such states, the use of appropriate multi-hadron operators is 
crucial.  Multi-hadron operators with significant coupling to the low-lying 
states of interest can be obtained by combining single-hadron operators
of various momenta.  The construction and testing of single-hadron
operators of definite momentum and their combinations into two-hadron
operators are the subject of this work.

Our approach to constructing single baryon operators of zero momentum
was previously described in Ref.~\cite{baryons2005A}.  A slightly
different method was reported in Ref.~\cite{baryons2005B}.
Our first study of the nucleon and $\Delta$ excitations in the quenched
approximation was presented in Ref.~\cite{baryon2007}, and nucleon results 
for two flavors of dynamical quarks appeared in Ref.~\cite{nucleon2009}.
A survey of excited-state energies in small volume for the isovector mesons
and kaons using $N_f=2+1$ dynamical quarks was given in 
Ref.~\cite{Morningstar:2010ae}, along with results for the $\Lambda, \Sigma, 
\Xi$ baryons.  To extend our efforts into larger volumes and towards $u,d$
quark masses yielding lighter pions, the issue of multi-hadron states 
was addressed in Ref.~\cite{Morningstar:2011ka}.  A new stochastic method 
of treating the low-lying modes of quark propagation which exploits 
Laplacian Heaviside quark-field smearing was presented in that work,
although the method was briefly introduced with preliminary testing in 
Refs.~\cite{Morningstar:2010ae,Foley:2010vv,Bulava:2010em}.
Other recent progress in calculating excited-state energies in lattice
QCD can be found in Refs.~\cite{Aoki:2009ix,Mahbub:2010jz,Engel:2010my,
Bulava:2010yg,Peardon:2011tt,Engel:2011aa,Edwards:2011jj,Mahbub:2010rm,
Aoki:2011yj,Mohler:2012na,Lang:2011mn,Mahbub:2012ri,Edwards:2012fx,Thomas:2011rh,
Moir:2013ub,Alexandrou:2013fsu}.

In this work, the approach of Ref.~\cite{baryons2005A} is extended
to meson operators of zero momentum and to both meson and baryon operators
having definite nonzero momentum.  A new glueball operator is also
introduced and tested.  To simplify our spectrum calculations as much
as possible and to increase the statistical precision of our results, we make 
use of single-hadron operators that transform irreducibly under all symmetries 
of a three-dimensional cubic lattice with periodic boundary 
conditions.  Our method of constructing such operators is described in detail
in this paper.  Spectrum results obtained using these operators will be presented
in later reports, although we present some testing of these operators here.  
Our operator design utilizes group-theoretical projections.
The point and space groups we use are well known, and the properties of their 
irreducible representations are widely available in the literature. However,
we collect together and present in this paper some of the specific group theory
details needed for our operator construction for the convenience of the reader
and as a record of our conventions and notation.

The Monte Carlo method commonly employed in QCD computations applies only to
space-time lattices of finite extent.  Hence, our goal is to obtain the
stationary-state energies of QCD in a cubic box using periodic boundary
conditions.  In such a cubic box, we no longer have full rotational symmetry, 
even in the continuous space-time limit.
The stationary states cannot be labelled by the usual spin-$J$ quantum
numbers. Instead, the stationary states in a box with periodic boundary 
conditions must be labelled by the irreducible representations (irreps) of 
the cubic space group, even in the continuum limit.  

This paper is organized as follows.  Our approach to building single-hadron 
operators of definite momentum is described in Sec.~\ref{sec:singlehadron}.  
We construct operators that transform irreducibly under all symmetries of a 
three-dimensional cubic lattice.  Monte Carlo calculations are required
to test our operator construction, and we present the implementation details 
of such computations in Sec.~\ref{sec:implement}.  Tests of our single-hadron 
operators using the stochastic LapH method on $24^3\times 128$ and 
$32^3\times 256$ anisotropic lattices with pion masses $m_\pi\approx 390$ and 
240~MeV are then presented in Sec.~\ref{sec:singletests}.  How we combine
these operators to form two-hadron operators is described in 
Sec.~\ref{sec:twohadron}, and initial tests are presented.  In
Sec.~\ref{sec:glueball}, a new glueball operator is introduced and we
demonstrate the feasibility of evaluating the mixing of this glueball operator 
with a quark-antiquark operator, an $\eta\eta$ operator, and multiple two-pion 
operators.  Concluding remarks are given in Sec.~\ref{sec:conclude}, along 
with our plans for future work.

\section{Single-hadron operators of definite momentum}
\label{sec:singlehadron}

We extract the finite-volume stationary-state energies of QCD from
matrices of temporal correlations 
$C_{ij}(t_F-t_0)=\langle 0\vert\,T\, O_i(t_F)\,\overline{O}_j(t_0)\,\vert 0\rangle$,
where $T$ denotes time-ordering, the source operators $\overline{O}_j(t_0)$ create 
the states of interest at an initial time $t_0$, and the sink operators $O_i(t_F)$ 
annihilate the states of interest at a later time $t_F$.  
The correlation functions $C_{ij}(t)$ can be expressed in terms of ``path" 
integrals over quark fields $\overline{\psi},\psi$ and gluon fields $U$
involving the QCD action having the form
\begin{equation}
   S[\overline{\psi},\psi,U] = \overline{\psi} K[U]\psi + S_G[U],
\end{equation}
where $K[U]$ is known as the Dirac matrix and $S_G[U]$ is the gauge-field
action.  We use an anisotropic space-time lattice in which the temporal spacing 
$a_t$ is smaller than the spacing $a_s$ in the three spatial dimensions.  Each
time slice is a three-dimensional cubic lattice.  We assume periodic boundary
conditions in all three spatial directions.  In order 
to estimate the path integrals using the Monte Carlo method, it is necessary to 
work in the imaginary time formalism.  This also has the advantage of converting 
oscillatory factors in real time into decaying exponentials.  Our Euclidean space-time 
conventions, including the Euclidean Dirac $\gamma$-matrices, are given 
in Ref.~\cite{baryons2005A}.  In the four-vector $x_\mu$ describing a given lattice
site, the $\mu=4$ component specifies the position in time, and the $\mu=1,2,3$ components
specify the position along the Cartesian $x,y,z$ spatial directions. As usual in lattice 
gauge theory, the gluon field is introduced using the parallel transporter $U_\mu(x)$ given
by the path-ordered exponential of the gauge field along a link in the $\mu$ direction
connecting neighboring sites of the lattice.  The Dirac spinor field $\psi^A_{a\alpha}(x)$ 
annihilates a quark and creates an antiquark at lattice site $x$, where $A$ refers to the 
quark flavor, $a$ refers to color, and $\alpha$ is the Dirac spin index, and the field 
$\overline{\psi}^A_{a\alpha}(x)$ annihilates an antiquark and creates a quark.  
Unlike in Minkowski space-time, $\psi$ and $\overline{\psi}$ must be treated as 
independent fields.  When used in path integrals, the link variables are $SU(3)$
matrices and $\overline{\psi},\psi$ are complex Grassmann fields.

Our hadron operators are constructed using spatially-smoothed link variables 
$\widetilde{U}_j(x)$ and spatially-smeared quark fields $\widetilde{\psi}(x)$.  
The spatial links are smeared using the stout-link procedure described in 
Ref.~\cite{Morningstar:2003gk}. Note that only spatial staples are used in the 
link smoothening; no temporal staples are used.  The smeared quark field for 
each quark flavor is defined by
\begin{equation}
\widetilde{\psi}_{a\alpha}(x) =
   {\cal S}_{ab}(x,y)\ \psi_{b\alpha}(y),
\end{equation}
where $x,y$ are lattice sites, $a,b$ are color indices, and $\alpha$ is a 
Dirac spin component. We use the Laplacian Heaviside (LapH) quark-field smearing 
scheme introduced in Ref.~\cite{distillation2009} and defined by
\begin{equation}
{\cal S} = 
 \Theta\left(\sigma_s^2+\widetilde{\Delta}\right),
\end{equation}
where $\widetilde{\Delta}$ is the three-dimensional gauge-covariant Laplacian
defined in terms of the stout-smeared gauge field $\widetilde{U}$, and $\sigma_s$
is the smearing cutoff parameter.  More details concerning this smearing
scheme are described in Ref.~\cite{Morningstar:2011ka}.

All of our single-hadron operators are assemblages of basic building blocks
which are gauge-covariantly-displaced, LapH-smeared quark fields:
\begin{equation}
 q^A_{a\alpha j}= D^{(j)}\widetilde{\psi}_{a\alpha}^{(A)},
 \qquad  \overline{q}^A_{a\alpha j} = \widetilde{\overline{\psi}}_{a\alpha}^{(A)}
  \gamma_4\, D^{(j)\dagger},
\label{eq:quarkdef}
\end{equation}
where $a$ is a color index, $\alpha$ is a Dirac spin component, $A$ is a quark flavor, 
$\gamma_4$ is the temporal Dirac $\gamma$-matrix, and $D^{(j)}$ is a gauge-covariant 
displacement of type $j$.  The displacement type is a sequence of $p$ spatial directions 
on the lattice $j=(j_1,j_2,\cdots,j_p)$.  This displacement can be trivial ($j=0$ meaning 
no displacement), a displacement in a given single spatial direction on the lattice by 
some number of links (typically two or three), or a combination of two or more spatial 
lattice directions.  If we define $d_r = \hat{j}_1+\hat{j}_2+\dots+\hat{j}_{r-1}$,
then the displacement $D^{(j)}$ is defined as a product of smeared link variables:
\begin{eqnarray}
 D^{(j)}(x,x^\prime) &=&
 \widetilde{U}_{j_1}(x)\ \widetilde{U}_{j_2}(x\!+\!d_2)
 \ \widetilde{U}_{j_3}(x\!+\!d_3)\dots \nonumber \\
&\times &  \widetilde{U}_{j_p}(x\!+\!d_p)
  \delta_{x^\prime,x+d_{p+1}}.
\end{eqnarray}
The use of $\gamma_4$ in Eq.~(\ref{eq:quarkdef}) is convenient for obtaining baryon 
correlation matrices that are Hermitian. To simplify notation, the Dirac spin component 
and the displacement type are sometimes combined into a single index in what follows.

We can simplify our spectrum calculations as much as possible by working with single-hadron 
operators that transform irreducibly under all symmetries of a three-dimensional cubic 
lattice of infinite extent or finite extent with periodic boundary conditions.
Such symmetries form the simple cubic space group known as $O_h^1$ in Sch\"onflies notation or 
Pm$\overline{3}$m in international notation.  This crystallographic space group is a 
semi-direct product of the abelian group of allowed translations on a simple cubic lattice 
and the orthogonal point group $O_h$.  For bosonic systems with zero strangeness, we 
add $G$-parity as a symmetry operation.  

\begin{figure}
\begin{center}
\includegraphics[width=50mm,bb= 0 40 329 299 ]{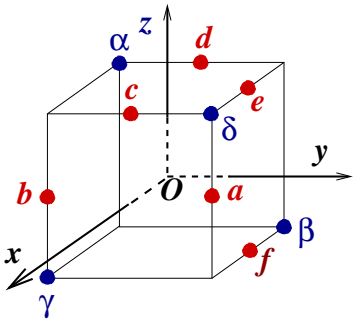}
\end{center}
\caption{ The axes $Ox$, $Oy$, $Oz$, $Oa$, $Ob$, $Oc$, $Od$, $Oe$,
 $Of$, $O\alpha$, $O\beta$, $O\gamma$, $O\delta$ corresponding to the
 group elements $C_{nj}$ of the octahedral group $O$, where $C_{nj}$
 denotes a proper rotation through angle $2\pi/n$ about axis $Oj$.
\label{fig:axes}}
\end{figure}

In order to construct such hadron operators, we first need to identify all of the 
symmetry operations and to determine how our basic building blocks transform under 
these operations.  An element of the space group $O_h^1$ is here denoted
by $(R,\bvec)$, where $R$ is a spatial rotation or reflection about the 
origin and is an element of the point group $O_h$, and $\bvec$ is an allowed 
spatial shift.  The group element $(R,\bvec)$ corresponds to the coordinate
transformation $\xvec\rightarrow R\xvec+\bvec$.  The covariantly-displaced quark
fields transform according to
\begin{eqnarray}
  U_{(R,\bvec)}\, q^A_{a\alpha\, j}(\xvec)\, U_{(R,\bvec)}^\dagger &=&
  S(R)^{-1}_{\alpha\beta}\  q^A_{a\beta\,Rj}(R\xvec\!+\!\bvec), \label{eq:transform}\\
  U_{(R,\bvec)}\, \overline{q}^A_{a\alpha\, j}(\xvec)\, U_{(R,\bvec)}^\dagger &=&
  \overline{q}^A_{a\beta\, Rj}(R\xvec\!+\!\bvec)\, S(R)_{\beta\alpha}, 
\end{eqnarray}
where $U_{(R,\bvec)}$ denotes the quantum field operator that effects the
transformation $(R,\bvec)$, and the transformation matrices for spatial 
inversion $I_s$ and proper rotations $C_{nj}$ through angle $2\pi/n$ about axis
$Oj$ are given by
\begin{eqnarray}
 S(C_{nj}) &=& \exp\Bigl(\textstyle\frac{1}{8}\omega_{\mu\nu}
 [\gamma_\mu,\gamma_\nu]  \Bigr),\\
S(I_s) &=& \gamma_4,
\end{eqnarray}
with $\omega_{kl}=-2\pi\varepsilon_{jkl}/n$ and $\omega_{4k}=\omega_{k4}=0$
($\omega_{\mu\nu}$ is an antisymmetric tensor which parametrizes rotations
and boosts). A rotation by $\pi/2$ about the $y$-axis is conventionally 
denoted by $C_{4y}$, and $C_{4z}$ denotes a rotation by $\pi/2$ about the 
$z$-axis.  These particular group elements are given by
\begin{equation}
 S(C_{4y})=\frac{1}{\sqrt{2}}(1+\gamma_1\gamma_3),\quad 
 S(C_{4z})=\frac{1}{\sqrt{2}}(1+\gamma_2\gamma_1).
\end{equation}
The allowed rotations on a three-dimensional spatially-isotropic cubic lattice
form the octahedral group $O$ which has 24 elements.  For the convenience of
the reader, the rotation axes corresponding to these group elements are shown 
in Fig.~\ref{fig:axes}.
Inclusion of spatial inversion $I_s$ yields the point group $O_h$ which has 
48 elements occurring in ten conjugacy classes. All elements of $O_h$ can be 
generated from appropriate products of only $C_{4y}$, $C_{4z}$, and $I_s$.
Under $G$-parity, our basic building blocks transform according to
\begin{eqnarray}
  U_G\ q^A_{a\alpha j}(x)\ U_G^\dagger &=&
  \overline{q}^B_{a\beta j}(x)\, (\gamma_2)_{\beta\alpha}\ G^{BA}, \\
  U_G\ \overline{q}^A_{a\alpha j}(x)\ U_G^\dagger &=&
  (\gamma_2)_{\alpha\beta}\  q^B_{a\beta j}(x)\ G^{BA}, 
\end{eqnarray}
using the Dirac-Pauli representation for the $\gamma$-matrices, and
where the only nonzero elements of the $G$ flavor matrix are 
$G^{ud}=-G^{du}=-G^{ss}=1$.

The construction of irreducible representations (irreps) of $O_h^1$ begins
with the irreps of the abelian subgroup of lattice translations.  These are 
characterized by a definite three-momentum $\pvec$ as allowed by the 
periodic boundary conditions.

Each of our meson operators which destroys a three-momentum $\pvec$ is a linear 
superposition of gauge-invariant quark-antiquark elemental operators of the form
\begin{equation}
 \Phi^{AB}_{\alpha\beta}(t)= 
\sum_{\bm{x}}e^{-i\pvec\cdot(\xvec+\frac{1}{2}(\bm{d}_\alpha+\bm{d}_\beta))}
  \delta_{ab}  \ \overline{q}^A_{a\alpha}(\bm{x},t)\ q^B_{b\beta}(\bm{x},t),
\label{eq:mesonelemental}
\end{equation}
where $q,\overline{q}$ are defined in Eq.~(\ref{eq:quarkdef}),
$\bm{d}_\alpha, \bm{d}_\beta$ are the spatial displacements of the
$\overline{q},q$ fields, respectively, from $\xvec$,
$A,B$ indicate flavor, and $\alpha,\beta$ are compound indices
incorporating both spin and quark-displacement types.  The phase factor
involving the quark-antiquark displacements is needed to ensure proper
transformation properties under $G$-parity for arbitrary displacement types.
The ``barred'' operators  which create a momentum $\pvec$ then take the form
\begin{equation}
 \overline{\Phi}_{\alpha\beta}^{AB}(t)=
 \sum_{\bm{x}} e^{i\pvec\cdot(\xvec+\frac{1}{2}(\bm{d}_\alpha+\bm{d}_\beta))}
   \delta_{ab}\ \overline{q}^B_{b\beta}(\bm{x},t)\ q^A_{a\alpha}(\bm{x},t).
\end{equation}
Each meson sink operator has the form
\begin{equation}
  M_{l}(t)= c^{(l)}_{\alpha\beta}\ \Phi^{AB}_{\alpha\beta}(t),
\label{eq:mesonstart}
\end{equation}
(or is a flavor combination of the above form),
where $l$ is a compound index comprised of a three-momentum $\pvec$, an
irreducible representation $\Lambda$ of the little group of $\pvec$ (discussed below), 
the row $\lambda$ of the irrep, total isospin $I$, isospin projection $I_3$,
strangeness $S$, and an identifier labeling the different operators in each 
symmetry channel.  Here, we focus on mesons containing only $u,d,s$ quarks.
The corresponding source operators are
\begin{equation}
  \overline{M}_{l}(t)= c^{(l)\ast}_{
 \alpha\beta}\ \overline{\Phi}^{AB}_{\alpha\beta}(t).
\end{equation}

\begin{table}
\caption[captab]{
The spatial arrangements of the quark-antiquark meson operators (left)
and the three-quark baryon operators (right). In the illustrations, the smeared quarks 
fields are depicted by solid circles, each hollow circle indicates a smeared ``barred'' 
antiquark field, the solid line segments indicate covariant displacements, and each 
hollow box indicates the location of a Levi-Civita color coupling.  For simplicity, 
all displacements have the same length in an operator.  
\label{tab:opforms}}
\begin{ruledtabular}
\begin{tabular}{cl}
 Meson configurations &  \qquad Baryon configurations\\ \hline
\raisebox{3mm}{
\begin{minipage}{1.6in} 
\raisebox{0mm}{\setlength{\unitlength}{1mm}
\thicklines
\begin{picture}(20,12)
\put(9,7){\circle{2}}
\put(11,7){\circle*{2.5}}
\put(3,2){single-site}
\end{picture}}\\[-2mm]
\raisebox{0mm}{\setlength{\unitlength}{1mm}
\thicklines
\begin{picture}(20,12)
\put(7,7){\circle{2}}
\put(13,7){\circle*{2.5}}
\put(8,7){\line(1,0){4}}
\put(-1,2){singly-displaced}
\end{picture}}  \\[0mm]
\raisebox{0mm}{\setlength{\unitlength}{1mm}
\thicklines
\begin{picture}(20,12)
\put(7,11){\circle{2}}
\put(13,5){\circle*{2.5}}
\put(12,5){\line(-1,0){5}}
\put(7,10){\line(0,-1){5}}
\put(-3,0){doubly-displaced-L}
\end{picture}} \\[1mm]
\raisebox{0mm}{\setlength{\unitlength}{1mm}
\thicklines
\begin{picture}(20,12)
\put(8,11){\circle{2}}
\put(14,11){\circle*{2.5}}
\put(8,4){\line(1,0){6}}
\put(14,4){\line(0,1){6}}
\put(8,4){\line(0,1){6}}
\put(-2,0){triply-displaced-U}
\end{picture}}\\[0mm]
\raisebox{0mm}{\setlength{\unitlength}{1mm}
\thicklines
\begin{picture}(20,15)
\put(7,5){\circle{2}}
\put(11,7){\circle*{2.5}}
\put(7,12){\line(1,0){8}}
\put(7,6){\line(0,1){6}}
\put(15,12){\line(-3,-4){3.0}}
\put(-3,0){triply-displaced-O}
\end{picture}}
\end{minipage}} &
\raisebox{-2mm}{
\begin{minipage}{1.6in}
\raisebox{0mm}{\setlength{\unitlength}{1mm}
\thicklines
\begin{picture}(16,10)
\put(5.25,4){\line(1,0){5.25}}
\put(5.25,9.5){\line(1,0){5.25}}
\put(5.25,4){\line(0,1){5.5}}
\put(10.5,4){\line(0,1){5.5}}
\put(7,6){\circle*{2}}
\put(9,6){\circle*{2}}
\put(8,8){\circle*{2}}
\put(1,0){single-site}
\end{picture}}  \\ 
\raisebox{0mm}{\setlength{\unitlength}{1mm}
\thicklines
\begin{picture}(23,10)
\put(5.25,3.5){\line(1,0){3.5}}
\put(5.25,3.5){\line(0,1){5.5}}
\put(5.25,9.0){\line(1,0){3.5}}
\put(8.75,3.5){\line(0,1){5.5}}
\put(7,5){\circle*{2}}
\put(7,7.3){\circle*{2}}
\put(14,6){\circle*{2}}
\put(8.75,6){\line(1,0){4.5}}
\put(-1,0){singly-displaced}
\end{picture}}   \\ 
\raisebox{0mm}{\setlength{\unitlength}{1mm}
\thicklines
\begin{picture}(26,8)
\put(10.4,3.5){\line(1,0){3}}
\put(10.4,3.5){\line(0,1){3}}
\put(10.4,6.5){\line(1,0){3}}
\put(13.4,3.5){\line(0,1){3}}
\put(12,5){\circle*{2}}
\put(6,5){\circle*{2}}
\put(18,5){\circle*{2}}
\put(6,5){\line(1,0){4.2}}
\put(18,5){\line(-1,0){4.4}}
\put(0,0){doubly-displaced-I}
\end{picture}} \\ 
\raisebox{0mm}{\setlength{\unitlength}{1mm}
\thicklines
\begin{picture}(20,13)
\put(6.5,3.5){\line(1,0){3}}
\put(6.5,3.5){\line(0,1){3}}
\put(6.5,6.5){\line(1,0){3}}
\put(9.5,3.5){\line(0,1){3}}
\put(8,5){\circle*{2}}
\put(8,11){\circle*{2}}
\put(14,5){\circle*{2}}
\put(14,5){\line(-1,0){4.4}}
\put(8,11){\line(0,-1){4.2}}
\put(-3,0){doubly-displaced-L}
\end{picture}}  \\ 
\raisebox{0mm}{\setlength{\unitlength}{1mm}
\thicklines
\begin{picture}(20,12)
\put(9,9){\line(1,0){2}}
\put(9,11){\line(1,0){2}}
\put(9,9){\line(0,1){2}}
\put(11,9){\line(0,1){2}}
\put(4,10){\circle*{2}}
\put(16,10){\circle*{2}}
\put(10,4){\circle*{2}}
\put(4,10){\line(1,0){5}}
\put(16,10){\line(-1,0){5}}
\put(10,4){\line(0,1){5}}
\put(-3,0){triply-displaced-T}
\end{picture}}    \\ 
\raisebox{0mm}{\setlength{\unitlength}{1mm}
\thicklines
\begin{picture}(20,12)
\put(9,9){\line(1,0){2}}
\put(9,11){\line(1,0){2}}
\put(9,9){\line(0,1){2}}
\put(11,9){\line(0,1){2}}
\put(6,6){\circle*{2}}
\put(16,10){\circle*{2}}
\put(10,4){\circle*{2}}
\put(9,9){\line(-1,-1){3.6}}
\put(16,10){\line(-1,0){5}}
\put(10,4){\line(0,1){5}}
\put(-3,0){triply-displaced-O}
\end{picture}} \vspace*{3mm}
\end{minipage} }
\end{tabular}
\end{ruledtabular}
\end{table}

Each of our baryon operators destroying a three-momentum $\pvec$ is a 
linear superposition of gauge-invariant elemental three-quark operators 
of the form
\begin{equation}
  \Phi^{ABC}_{\alpha\beta\gamma}(\pvec,t)= 
\sum_{\bm{x}} e^{-i\pvec\cdot\xvec}\varepsilon_{abc}
\, q^A_{a\alpha}(\bm{x},t)
\, q^B_{b\beta}(\bm{x},t)
\, q^C_{c\gamma}(\bm{x},t).
\label{eq:baryonelemental}
\end{equation}
The ``barred'' three-quark elemental operators which create a 
momentum $\pvec$ have the form
\begin{equation}
  \overline{\Phi}_{\alpha\beta\gamma}^{ABC}(\pvec,t)= 
 \sum_{\bm{x}} e^{i\pvec\cdot\xvec}\varepsilon_{abc}
\ \overline{q}^C_{c\gamma}(\bm{x},t)
\ \overline{q}^B_{b\beta}(\bm{x},t)
\ \overline{q}^A_{a\alpha}(\bm{x},t). 
\end{equation}
Our baryon sink operators, being linear superpositions of the three-quark
elemental operators, have the form
\begin{equation}
  B_{l}(t)= c^{(l)}_{
 \alpha\beta\gamma}\ \Phi^{ABC}_{\alpha\beta\gamma}(t),
\label{eq:baryonstart}
\end{equation}
where again, the $l$ label includes the momentum $\pvec$, the little
group irrep $\Lambda$, the row $\lambda$ of the irrep, isospin $I$,
isospin projection $I_3$, strangeness $S$, and an identifier
specifying the different operators in each symmetry channel. 
We focus on baryons containing only $u,d,s$ quarks.
The corresponding source operators are
\begin{equation}
  \overline{B}_{l}(t)= c^{(l)\ast}_{
 \alpha\beta\gamma}\ \overline{\Phi}^{ABC}_{\alpha\beta\gamma}(t).
\end{equation}

In order to build up the necessary orbital and radial structures expected
in the hadron excitations, we use a variety of spatially-extended configurations
for our hadron operators, as shown in Table~\ref{tab:opforms}.  First,
consider the zero-momentum operators.  The simplest meson operators combine
the quark and antiquark fields at the same lattice site.  We refer to these as
single-site (SS) operators.  In the singly-displaced (SD) meson operators, the quark
is displaced from the antiquark along a direction parallel to one of the axes of 
the lattice.  If
the quark is covariantly displaced from the antiquark along an L-shaped path,
we refer to this as a doubly-displaced-L (DDL) operator.  Displacement of the quark
along a U-shaped path or in three orthogonal directions from the antiquark leads to 
triply-displaced-U (TDU) and triply-displaced-O (TDO) meson operators, respectively, as shown 
in Table~\ref{tab:opforms}.  The simplest baryon operators combine the three
quarks at a single lattice site. In the singly-displaced baryons, one of the quarks
is displaced from the other two along a direction parallel to one of the axes
of the lattice.  Displacement
of two quarks away from the site of the Levi-Civita coupling leads to
doubly-displaced-I (DDI) and doubly-displaced-L baryon operators of zero momentum.
All three quarks can be displaced from the color-coupling site, producing
triply-displaced-T (TDT) and triply-displaced-O baryon operators of zero momentum,
as illustrated in Table ~\ref{tab:opforms}.
For simplicity, all displacement lengths along each of the different directions are
taken to be the same in any given operator. For mesons, we use a length of
$3a_s$, and for baryons, the length is $2a_s$, as will be discussed later.

For nonzero momenta, we restrict our attention to on-axis momenta, such as in the
$\pm\ex,\ \pm\ey,\ \pm\ez$ directions, to momenta in a planar-diagonal direction,
such as $\pm\ex\pm\ey,\ \pm\ex\pm\ez,\ \pm\ey\pm\ez$, and momenta in a 
cubic-diagonal direction, such as $\pm\ex\pm\ey\pm\ez$.  We expect that the above
momentum directions are sufficient for studying the stationary-states in the 
range of energies of interest to us.  For on-axis momenta, we construct
single-site meson and baryon operators, longitudinally-singly-displaced (LSD) operators
in which one quark is displaced along the direction of the momentum, and
transverse-singly-displaced (TSD) operators, in which one quark is displaced along
a direction of the lattice transverse to the momentum.  For planar-diagonal momenta,
we use single-site meson and baryon operators, transverse-singly-displaced operators 
in which one quark is displaced along the direction of the lattice that is perpendicular 
to the plane containing the momentum direction, and planar-singly-displaced (PSD) operators 
in which one quark is displaced along one of the two directions of the lattice coinciding
with the nonzero components of the momentum.  For cubic-diagonal momenta, we use 
single-site and singly-displaced (SD) configurations for both baryons and mesons.  
For such momenta, displacements along the lattice axes are neither entirely transverse 
nor entirely longitudinal to the momentum.  For mesons, we also use triply-displaced-O 
(TDO) operators for the on-axis, planar-diagonal, and cubic-diagonal momenta.

\begin{table}
\caption{Our choices for the reference momentum $\pvec_{\rm ref}$ directions and 
the reference rotations $R_{\rm ref}$ for each momentum $\pvec$ direction that we use.
\label{tab:refmoms}}
\begin{ruledtabular}
\begin{tabular}{ccc}
$\pvec_{\rm ref}$ direction & $\pvec$ direction & $R_{\rm ref}^{\pvec}$\\ \hline
$(0,0,1)$ &  $(~0,~0,-1)$ & $C_{2x}$  \\
          &  $(~1,~0,~0)$ & $C_{4y}$           \\
          &  $(-1,~0,~0)$ & $C_{4y}^{-1}$  \\
          &  $(~0,-1,~0)$ & $C_{4x}$           \\
          &  $(~0,~1,~0)$ & $C_{4x}^{-1}$  \\ \hline
$(0,1,1)$ & $(0,-1,-1)$ & $C_{2x}$         \\
          & $(0,~1,-1)$ & $C_{4x}^{-1}$     \\
          & $(0,-1,1)$  & $C_{4x}$         \\
          & $(~1,0,~1)$ & $C_{4z}^{-1}$      \\
          & $(-1,0,-1)$ & $C_{2b}=C_{2x}C_{4z}$ \\
          & $(~1,0,-1)$ & $C_{2a}=C_{2y}C_{4z}$ \\
          & $(-1,0,~1)$ & $C_{4z}$          \\
          & $(~1,~1,0)$ & $C_{4y}$           \\
          & $(-1,-1,0)$ & $C_{2d}=C_{2z}  C_{4y}$   \\
          & $(~1,-1,0)$ & $C_{2c}=C_{4y}  C_{2z}$ \\
          & $(-1,~1,0)$ & $C_{4y}^{-1}$     \\ \hline
$(~1,~1,~1)$ & $(~1,~1,-1)$& $C_{4y}$  \\
 & $(~1,-1,~1)$& $C_{4x}$  \\
 & $(~1,-1,-1)$& $C_{2x}$  \\
 & $(-1,~1,~1)$& $C_{4z}$  \\  
 & $(-1,1,-1)$ & $C_{2y}$  \\
 & $(-1,-1,~1)$& $C_{2z}$  \\
 & $(-1,-1,-1)$& $C_{2d}=C_{2z}  C_{4y}$  \\
\end{tabular}
\end{ruledtabular}
\end{table}

For a given flavor structure, the next step in our single-hadron operator construction 
is to find coefficients in Eqs.~(\ref{eq:mesonstart}) and (\ref{eq:baryonstart}) that 
produce operators which transform irreducibly under all symmetries of the 
three-dimensional cubic lattice.  First, for each class of momenta, such as on-axis or 
planar-diagonal, we choose one representative reference momentum direction 
$\pvec_{\rm ref}$.  We then find coefficients corresponding to operators that
transform irreducibly under the little group of $\pvec_{\rm ref}$.  Recall that the 
little group of $\pvec_{\rm ref}$ is the subset of symmetry operations that leave the 
reference momentum $\pvec_{\rm ref}$ invariant.  Next, for each momentum direction $\pvec$
in a class of momenta, we select one reference rotation $R_{\rm ref}^{\pvec}$ that 
transforms $\pvec_{\rm ref}$ into $\pvec$.  As long as the
selected group element transforms $\pvec_{\rm ref}$ into $\pvec$, it does not matter 
which group element is chosen, but a choice must be made and clearly specified.
Hadron operators having a momentum in the direction of $\pvec$ are then obtained
by applying the reference rotation to the operators constructed using the momentum
in the direction of $\pvec_{\rm ref}$.  Our choices of reference momenta directions
and reference rotations are listed in Table~\ref{tab:refmoms}.

Our choices of reference directions and rotations, as well as our
choices of the irreducible representation matrices, described later, are 
dictated mainly by simplicity.  An alternate approach would be to choose the 
$z$-direction as the single reference direction, obtain all other momenta 
using a rotation defined by the Jacob-Wick convention, and use
irreducible representation matrices corresponding to helicity states.
Since our procedure for combining the single-hadron operators into
multi-hadron operators is automated using Maple, we found that there was
no great advantage in using the standard Jacob-Wick convention with 
helicities.

The little groups associated with our choices of reference momentum directions
are listed in Tables~\ref{tab:Ohelements}, \ref{tab:C4velements}, 
\ref{tab:C2velements}, and \ref{tab:C3velements}.  To describe both mesons
and baryons, we need the single-valued and double-valued (spinorial)
irreps of these groups.  The double-valued representations of a 
group ${\cal G}$ are constructed by extending the group elements to form the 
so-called ``double group'' ${\cal G}^D$.  This is done by introducing a new 
generator, denoted by $\overline{E}$, which represents a rotation by $2\pi$ 
about any axis.  For each element $G$ of the original group, the double
group contains another element $\overline{G}=\overline{E}G$.
For the convenience of the reader, the elements of the double groups
associated with our choices of reference momentum directions are explicitly
listed in Tables~\ref{tab:Ohelements}, \ref{tab:C4velements}, 
\ref{tab:C2velements}, and \ref{tab:C3velements}, grouped into 
their conjugacy classes. 

\begin{table}
\caption{The little group corresponding to reference momentum direction
 $(0,0,0)$ is $O_h$.  The elements of the double group $O_h^D$ are listed
 below, grouped into conjugacy classes.  $E$ is the identity element, 
 $\overline{E}$ represents a rotation
 by $2\pi$ about any axis, and $\overline{G}=\overline{E}G$ for each
 element $G$ in $O_h$.  Spatial inversion $I_s$ in $O_h^D$ satisfies 
 $I_s^2=E$ and $I_s^{-1}=I_s$.  Conjugacy classes ${\cal C}_9$ through
 ${\cal C}_{16}$ are not listed below.  The elements of class ${\cal C}_{n+8}$
 are obtained by multiplying each of the elements in class ${\cal C}_n$
 by $I_s$.
\label{tab:Ohelements}}
\begin{ruledtabular}
\begin{tabular}{l}
 ${\cal C}_1 = \{ E \}$ \\
 ${\cal C}_2 = \{ C_{3\alpha}, C_{3\beta}, C_{3\gamma}, C_{3\delta},
                   C^{-1}_{3\alpha}, C^{-1}_{3\beta}, C^{-1}_{3\gamma}, 
                   C^{-1}_{3\delta} \}$\\
 ${\cal C}_3 = \{ C_{2x}, C_{2y}, C_{2z}, \overline{C}_{2x}, 
                   \overline{C}_{2y}, \overline{C}_{2z} \}$\\
 ${\cal C}_4 = \{ C_{4x}, C_{4y}, C_{4z},
                   C^{-1}_{4x}, C^{-1}_{4y}, C^{-1}_{4z} \}$\\
 ${\cal C}_5 = \{ C_{2a}, C_{2b}, C_{2c}, C_{2d}, C_{2e}, C_{2f},$\\
              \qquad\qquad$ \overline{C}_{2a}, \overline{C}_{2b},
                   \overline{C}_{2c}, \overline{C}_{2d},
                   \overline{C}_{2e}, \overline{C}_{2f} \} $\\
 ${\cal C}_6 = \{ \overline{E} \} $\\
 ${\cal C}_7 = \{ \overline{C}_{3\alpha}, \overline{C}_{3\beta}, 
                   \overline{C}_{3\gamma}, \overline{C}_{3\delta},
                   \overline{C}^{-1}_{3\alpha}, \overline{C}^{-1}_{3\beta},
                   \overline{C}^{-1}_{3\gamma}, \overline{C}^{-1}_{3\delta} \}$\\
 ${\cal C}_8 = \{ \overline{C}_{4x}, \overline{C}_{4y}, \overline{C}_{4z},
                   \overline{C}^{-1}_{4x}, \overline{C}^{-1}_{4y}, 
                   \overline{C}^{-1}_{4z} \}$
\end{tabular}
\end{ruledtabular}
\end{table}

\begin{table}
\caption{The little group corresponding to reference momentum direction
 $(0,0,1)$ is $C_{4v}$.  The elements of the double group $C_{4v}^D$ for this
 reference momentum direction are listed
 below, grouped into conjugacy classes.  $E$ is the identity element, 
 $\overline{E}$ represents a rotation by $2\pi$ about any axis.
\label{tab:C4velements}}
\begin{ruledtabular}
\begin{tabular}{l}
$\mathcal{C}_{1}=  \{  E  \} $ \\
$\mathcal{C}_{2}=  \{ C_{2 z}, \overline{C}_{2 z}  \} $ \\
$\mathcal{C}_{3}=  \{ C_{4 z}, C_{4z}^{-1}  \} $ \\
$\mathcal{C}_{4}=  \{ I_{s} C_{2 x}, I_{s} C_{2 y}, I_{s} 
  \overline{C}_{2 x}, I_{s} \overline{C}_{2 y} \} $  \\
$\mathcal{C}_{5}=  \{ I_{s} C_{2 a}, I_{s} C_{2 b}, I_{s} 
  \overline{C}_{2 a}, I_{s} \overline{C}_{2 b}  \} $ \\
$\mathcal{C}_{6}=  \{ \overline{E}  \} $ \\
$\mathcal{C}_{7}=  \{ \overline{C}_{4 z}, \overline{C}_{4z}^{-1}  \} $
\end{tabular}
\end{ruledtabular}
\end{table}

\begin{table}
\caption{The little group corresponding to reference momentum direction
 $(0,1,1)$ is $C_{2v}$.  The elements of the double group $C_{2v}^D$ for this
 reference momentum direction are listed
 below, grouped into conjugacy classes. 
\label{tab:C2velements}}
\begin{ruledtabular}
\begin{tabular}{l}
$\mathcal{C}_{1} =  \{E  \} $ \\
$\mathcal{C}_{2} =  \{ C_{2e}, \overline{C}_{2 e}  \} $ \\
$\mathcal{C}_{3} =  \{ I_{s} C_{2f}, I_{s} \overline{C}_{2 f}  \}$ \\
$\mathcal{C}_{4} =  \{ I_{s} C_{2x}, I_{s} \overline{C}_{2 x}  \}$ \\
$\mathcal{C}_{5} =  \{ \overline{E}  \} $
\end{tabular}
\end{ruledtabular}
\end{table}

\begin{table}
\caption{The little group corresponding to reference momentum direction
 $(1,1,1)$ is $C_{3v}$.  The elements of the double group $C_{3v}^D$ are listed
 below, grouped into conjugacy classes. 
\label{tab:C3velements}}
\begin{ruledtabular}
\begin{tabular}{l}
$\mathcal{C}_{1} = \{E  \} $ \\
$\mathcal{C}_{2} = \{ C_{3 \delta}, C_{3 \delta}^{-1}  \} $ \\
$\mathcal{C}_{3} = \{ I_{s} C_{2 b}, I_{s} C_{2 d}, I_{s} C_{2 f}  \} $\\
$\mathcal{C}_{4} = \{ \overline{E}  \} $ \\
$\mathcal{C}_{5} = \{ \overline{C}_{3 \delta}, \overline{C}_{3 \delta}^{-1}  \}$ \\
$\mathcal{C}_{6} = \{ I_{s} \overline{C}_{2 b}, I_{s} \overline{C}_{2 d}, I_{s}
 \overline{C}_{2 f}  \}$
\end{tabular}
\end{ruledtabular}
\end{table}

The irreducible representations of these little groups and their characters
are listed in Tables~\ref{tab:Ohcharacters}, \ref{tab:C4vcharacters}, 
\ref{tab:C2vcharacters}, and \ref{tab:C3vcharacters}.  One-dimensional
single-valued irreps are labelled by $A$ or $B$, two-dimensional irreps are
denoted by $E$, and three-dimensional irreps are labelled by $T$.  
One-dimensional double-valued irreps are denoted by $F$, two-dimensional
spinor irreps are denoted by $G$, and four-dimensional irreps are indicated
by $H$.  A subscript $g$ indicates an even-parity irrep, whereas a
subscript $u$ indicates an odd-parity irrep.  For mesons which are
eigenstates of $G$-parity, a ``$+$" superscript indicates an irrep describing
states even under $G$-parity, and a ``$-$" superscript indicates an irrep
associated with states odd under $G$-parity.  Our notation differs
from that of Refs.~\cite{Moore:2005dw,Moore:2006ng}.

Our method of constructing the hadron operators that transform irreducibly
under each little group makes use of group-theoretical projections and is
described in detail in Ref.~\cite{baryons2005A}.  The first step in the method
is to identify a basis of hadron elemental operators $\Phi_i(t)$ at a single 
time $t$ that transform into one another under the elements of the little
group ${\cal G}$.  The key formula in obtaining the linear combinations $O_i$ of 
these basis operators that transform irreducibly under ${\cal G}$ is
\begin{equation}
  O_{i}^{\Lambda\lambda}(t)
 = \frac{d_\Lambda}{g_{{\cal G}^D}}\sum_{R\in {\cal G}^D} 
  \Gamma^{(\Lambda)}_{\lambda\lambda}(R)\ U_R\ \Phi_i(t)\ U_R^\dagger,
\label{eq:project}
\end{equation}
where ${\cal G}^D$ is the double group of ${\cal G}$, $R$ denotes an element of 
${\cal G}^D$, $g_{{\cal G}^D}$ is the number of elements in ${\cal G}^D$, 
$d_\Lambda$ is the dimension of the $\Lambda$ irreducible representation,
and $\Gamma^\Lambda(R)$ is the matrix for element $R$ in irrep $\Lambda$.

To carry out the projections in Eq.~(\ref{eq:project}), explicit 
representation matrices $\Gamma$ are needed for each group element, and a 
representation for the Dirac $\gamma$-matrices must be selected.  We use 
the Dirac-Pauli representation for the $\gamma$-matrices, as described
in Ref.~\cite{baryons2005A}.
Often, helicity states are used for moving hadrons in continuous space-time.
We could find no great advantage in using a helicity representation
for our subsequent lattice calculations involving multi-hadron operators.
The use of helicity states does not result in any reduction in
the number of irrep rows that must be evaluated for the single moving
hadron states. All of our group theory manipulations are implemented 
using Maple, so the computational effort is independent of our choices of 
representation matrices.  Instead, we chose the simplest possible matrices
for our irreducible representations.
Our choices of representation matrices
are summarized in Tables~\ref{tab:ODreps}, \ref{tab:C4v_irrep_matrices},
\ref{tab:C2v_irrep_matrices}, and \ref{tab:C3v_irrep_matrices}.
The representation matrices for all group elements can be obtained
by suitable multiplications of the matrices shown in these tables.

\begin{table}
\caption[tabtwo]{Characters $\chi^\Lambda$ of the single-valued and double-valued 
 irreducible representations $\Lambda$ of the group $O_h$. Only the even-parity 
 irreps (subscript $g$) and classes ${\cal C}_1$ to ${\cal C}_8$ are shown below.
 For the even-parity irreps, $\chi_{n+8}^\Lambda=\chi_n^\Lambda$, where 
 $\chi_n^\Lambda$ denotes the character of $\Lambda$ for all group elements
 in class ${\cal C}_n$.  For the odd-parity irreps (subscript $u$ instead of $g$),
 $\chi^{\Lambda_u}_n=\chi^{\Lambda_g}_n$ for $n=1\dots 8$, and
 $\chi^{\Lambda_u}_n=-\chi^{\Lambda_g}_n$ for $n=9\dots 16$. 
\label{tab:Ohcharacters}}
\begin{ruledtabular}
\begin{tabular}{l|rrrrrrrr}
  $\Lambda$  & $\chi^\Lambda_1$ & $\chi^\Lambda_2$ & $\chi^\Lambda_3$ 
  & $\chi^\Lambda_4$  & $\chi^\Lambda_5$ & $\chi^\Lambda_6$ 
& $\chi^\Lambda_7$ & $\chi^\Lambda_8$ \\ \hline
  $A_{1g}$ & 1 & 1    & 1    & 1    & 1    & 1& 1   & 1   \\
  $A_{2g}$ & 1 & 1    & 1    & $-1$ & $-1$ & 1& 1   & $-1$\\
  $E_g$    & 2 & $-1$ & 2    & 0    & 0    & 2& $-1$& 0   \\
  $T_{1g}$ & 3 & 0    & $-1$ & 1    & $-1$ & 3& 0   & 1   \\
  $T_{2g}$ & 3 & 0    & $-1$ & $-1$ & 1    & 3& 0   & $-1$\\ 
  $G_{1g}$ & 2 & 1    & 0 & $\sqrt{2}$  & 0 & $-2$ & $-1$ & $-\sqrt{2}$   \\
  $G_{2g}$ & 2 & 1    & 0 & $-\sqrt{2}$ & 0 & $-2$ & $-1$ & $\sqrt{2}$    \\
  $H_g$   & 4 & $-1$ & 0 & 0           & 0 & $-4$ & $1$  & 0 
\end{tabular}
\end{ruledtabular}
\end{table}

\begin{table}
\caption{Characters $\chi^\Lambda$ for the single-valued and double-valued 
 irreps $\Lambda$ of the group $C_{4 v}$. 
\label{tab:C4vcharacters}}
\begin{ruledtabular}
\begin{tabular}{c|rrrrrrr}
$\Lambda$ & $\chi^\Lambda_{1}$ & $\chi^\Lambda_{2}$ & $\chi^\Lambda_{3}$ 
& $ \chi^\Lambda_{4}$ & $\chi^\Lambda_{5}$ & $\chi^\Lambda_{6}$ 
& $\chi^\Lambda_{7}$ \\ \hline
$A_{1}$ & $1$ & $1$ & $1$  &  $1$  & $1$  & $1$ & $ 1$\\
$A_{2}$ & $1$ & $1$ & $1$  &  $-1$ & $-1$ & $1$ & $ 1$\\
$B_{1}$ & $1$ & $1$ & $-1$ &  $1$  & $-1$ & $1$ & $-1$\\
$B_{2}$ & $1$ & $1$ & $-1$ &  $-1$ & $1$  & $1$ & $-1$\\
$E$     & $2$ & $-2$& $0$  &  $0$  & $0$  & $2$ & $ 0$\\
$G_{1}$ & $2$ & $0$ & $\sqrt{2}$ & $0$ & $0$ & $-2$ & $-\sqrt{2}$ \\
$G_{2}$ & $2$ & $0$ & $-\sqrt{2}$ & $0$ & $0$ & $-2$ & $\sqrt{2}$
\end{tabular}
\end{ruledtabular}
\end{table}

\begin{table}
\caption{Characters $\chi^\Lambda$ for the single-valued and double-valued 
 irreps $\Lambda$ of the group $C_{2v}$.}
\label{tab:C2vcharacters}
\begin{ruledtabular}
\begin{tabular}{c|rrrrr}
$\Lambda$ & $\chi^\Lambda_{1}$ & $\chi^\Lambda_{2}$ & $\chi^\Lambda_{3}$ 
& $\chi^\Lambda_{4}$ & $\chi^\Lambda_{5}$  \\ \hline
 $A_{1}$  & 1 & 1 & 1 & 1 & 1 \\
 $A_{2}$  & 1 & 1 & $-1$ & $-1$ & 1 \\
 $B_{1}$  & 1 & $-1$ & $1$ & $-1$ & 1  \\
 $B_{2}$  & 1 & $-1$ & $-1$ & $1$ & 1 \\
 $G$      & 2 & 0 & 0 & 0 & $-2$ 
\end{tabular}
\end{ruledtabular}
\end{table}

\begin{table}
\caption{Characters $\chi^\Lambda$ for the single-valued and double-valued
 irreps $\Lambda$ of the group $C_{3v}$.}
\label{tab:C3vcharacters}
\begin{ruledtabular}
\begin{tabular}{c|rrrrrr}
$\Lambda$ & $\chi^\Lambda_{1}$ & $\chi^\Lambda_{2}$ & $\chi^\Lambda_{3}$ & 
$\chi^\Lambda_{4}$ & $\chi^\Lambda_{5}$ & $\chi^\Lambda_{6}$ \\ \hline
 $A_{1}$  & 1 &  1 &  1 & 1 & 1 &  1  \\
 $A_{2}$  & 1 &  1 & $-1$ & 1 & 1 & $-1$  \\
 $E$      & 2 & $-1$ &  0 & 2 & $-1$ & 0  \\
 $F_{1}$  & 1 & $-1$ & $i$ & $-1$ & 1 & $-i$ \\
 $F_{2}$  & 1 & $-1$ & $-i$ & $-1$ & 1 & $i$ \\
 $G$      & 2 &  1 & 0 & $-2$ & $-1$ & 0 
\end{tabular}
\end{ruledtabular}
\end{table}

\begin{table}
\caption[tabODrep]{Our choices for the representation matrices $\Gamma$ of the 
 single-valued and double-valued irreps $\Lambda$ of the group $O_h$ for zero momentum 
 operators.  Only the even-parity irreps (subscript $g$) are shown below.  The matrices 
 for $C_{4y}$ and $C_{4z}$ for the odd-parity irreps (subscript $u$) are the same.
 Spatial inversion $I_s$ is the third generator.  $\Gamma^{(\Lambda)}(I_s)$ is
 the identity matrix for the even-parity irreps, and minus one times the identity
 for the odd-parity irreps. The matrices for all other 
 group elements can be obtained from appropriate multiplications of the matrices below
 and the matrix for $I_s$.
 \label{tab:ODreps}}
\begin{ruledtabular}
\begin{tabular}{ccc}
 $\Lambda$ & $\Gamma^{(\Lambda)}(C_{4y})$
 & $\Gamma^{(\Lambda)}(C_{4z})$ 
 \\ \hline
$A_{1g}$ \rule[-0ex]{0pt}{2ex} &
$[1]$ & $[1]$  \\ 
$A_{2g}$ \rule[-0ex]{0pt}{2ex} & $[-1]$ & $[-1]$  \\ 
$E_g$ \rule[-3ex]{0pt}{6ex} &
$\displaystyle\frac{1}{2}\left[\begin{array}{rr}
 1 & \sqrt{3} \\
 \sqrt{3} & -1
\end{array}\right] $ &
$ \left[\begin{array}{rr}
 -1 & 0 \\ 0 & 1 
\end{array}\right] $ \\
$T_{1g}$ \rule[-3ex]{0pt}{9ex} &
$\left[\begin{array}{rrr}
   0&0&1 \\ 0&1&0 \\ -1&0&0 \end{array}
 \right]$ &
$\left[\begin{array}{rrr}
 0&-1&0 \\ 1&0&0 \\ 0&0&1   \end{array}
 \right]$ \\ 
$T_{2g}$ \rule[-3ex]{0pt}{9ex} &
$\left[\begin{array}{rrr}
  0&0&-1 \\ 0&-1&0 \\ 1&0&0    \end{array}
 \right]$ &
$\left[\begin{array}{rrr}
0&1&0 \\ -1&0&0 \\ 0&0&-1      \end{array}
 \right]$ \\ 
$G_{1g}$  & \rule[-3ex]{0mm}{8ex}
$\displaystyle\frac{1}{\sqrt{2}}\!\left[\begin{array}{rr}
 1 & -1 \\ 1 &  1
\end{array}\right]$ &
$\displaystyle\frac{1}{\sqrt{2}}\!\left[\begin{array}{cc}
  1\!-\!i & 0 \\ 0 & 1\!+\!i 
\end{array}\right] $ \\
$G_{2g}$   &\rule[-3ex]{0mm}{8ex}
$\displaystyle\frac{-1}{\sqrt{2}}\!\left[\begin{array}{rr}
 1 & -1 \\ 1 &  1
\end{array}\right]$ &
 $\displaystyle\frac{-1}{\sqrt{2}}\!\left[\begin{array}{cc}
  1\!-\!i & 0 \\ 0 & 1\!+\!i 
\end{array}\right]$ \\
$H_g$  &\rule[-7ex]{0mm}{15ex}
$\displaystyle\!\!\frac{1}{2\sqrt{2}}\!\!\left[\begin{array}{rrrr}
  \!1 & \!\!-\sqrt{3} & \sqrt{3} & -1 \\
  \!\!\sqrt{3} & -1 & -1 &  \sqrt{3} \\
  \!\!\sqrt{3} &  1 & -1 & \!\!-\sqrt{3} \\
  1 &  \sqrt{3} & \sqrt{3} &  1 \end{array}\right] $
& $\displaystyle\!\!\frac{1}{\sqrt{2}}\!\!\left[\begin{array}{cccc}
   \!\!-1\!-\!i\!\! & 0 & 0 & 0 \\
   0 & \!\!1\!-\!i\!\! & 0 & 0 \\
   0 & 0 & \!\!1\!+\!i\!\! & 0 \\
   0 & 0 & 0 & \!\!-1\!+\!i\!\! \end{array}\right]$ 
\end{tabular}
\end{ruledtabular}
\end{table}

\begin{table}
\caption[C4v]{
 Our choices for the representation matrices $\Gamma$ of the single-valued and
 double-valued irreps $\Lambda$ of the little group $C_{4v}$ for momentum in the direction
 $(0,0,1)$.  The matrices for 
 all other group elements can be obtained from appropriate multiplications of the matrices below.
\label{tab:C4v_irrep_matrices}}
\begin{ruledtabular}
\begin{tabular}{ccc}
 $\Lambda$ & $\Gamma^{\left( \Lambda \right)} \left( C_{4 z} \right) $ & 
 $ \Gamma^{\left( \Lambda \right) } \left( I_{s} C_{2 y} \right) $ \\
\hline
$ A_{1} \rule[-0mm]{0pt}{1mm}$ & $ \left[  1 \right] $  &  $ \left[ 1 \right] $  \\
$ A_{2} \rule[-0mm]{0pt}{1mm}$ & $ \left[  1 \right] $  &  $ \left[ -1 \right] $  \\
$ B_{1} \rule[-0mm]{0pt}{1mm}$ & $ \left[  -1 \right] $  &  $ \left[ 1 \right] $  \\
$ B_{2} \rule[-0mm]{0pt}{1mm}$ & $ \left[  -1 \right] $  &  $ \left[ -1 \right] $  \\
$ E \rule[-3ex]{0pt}{8ex}$ & $ \left[ \begin{array}{cc} 0 & -1 \\ 1 & 0  \end{array} 
\right] $ &  $ \left[
 \begin{array}{cc} 1 & 0 \\ 0 & -1 \end{array} \right] $ \\
$ G_{1} \rule[-3ex]{0pt}{8ex}$ & $ \displaystyle\frac{1} {\sqrt{2}} \left[ \begin{array}{cc} 1-i & 0 
\\ 0 & 1+i \end{array} \right] $ 
& $ \left[ \begin{array}{cc} 0 & -1 \\ 1 & 0 \end{array} \right] $ \\
$ G_{2} \rule[-3ex]{0pt}{8ex}$ & $ \displaystyle\frac{-1} {\sqrt{2}} \left[ \begin{array}{cc} 1-i & 0 
\\ 0 & 1+i \end{array} \right] $ 
& $ \left[ \begin{array}{cc} 0 & -1 \\ 1 & 0 \end{array} \right] $ \\
\end{tabular}
\end{ruledtabular}
\end{table}

\begin{table}
\caption[C2v]{
  Our choices for the representation matrices $\Gamma$ of the single-valued and
  double-valued irreps $\Lambda$ of the little group $C_{2 v}$ for momentum in the 
  direction $(0,1,1)$. The matrices for all other group elements can be obtained 
  from appropriate multiplications of the matrices below.  $C_{2e}=C_{2z}C_{4x}$ 
  is a rotation about $(0,1,1)$, and $C_{2f}=C_{2y}C_{4x}$ is a rotation 
  about $(0,1,-1)$.
\label{tab:C2v_irrep_matrices}}
\begin{ruledtabular}
\begin{tabular}{ccc}
 $\Lambda$ & $ \Gamma^{\left( \Lambda \right) } \left( C_{2e} \right) $ 
 & $ \Gamma^{\left( \Lambda \right) } \left( I_{s} C_{2 f} \right) $ \\
\hline
$A_{1}\rule[-0mm]{0pt}{1mm}$ & $\left[ 1 \right]$ & $\left[1 \right] $ \\
$A_{2}\rule[-0mm]{0pt}{1mm}$ & $\left[ 1 \right]$ & $\left[-1 \right] $ \\
$B_{1}\rule[-0mm]{0pt}{1mm}$ & $\left[ -1 \right]$ & $\left[ 1 \right] $ \\
$B_{2}\rule[-0mm]{0pt}{1mm}$ & $\left[ -1 \right]$ & $\left[ -1 \right] $ \\
$G\rule[-3ex]{0pt}{7ex}$ & $ \displaystyle\frac{1} {\sqrt{2}} \left[  
\begin{array}{cc} -i & -1 \\ 
1 & i \end{array} \right] $  & 
 $ \displaystyle\frac{1} {\sqrt{2}} \left[  \begin{array}{cc} i & -1 \\ 
1 & -i \end{array} \right] $ 
\end{tabular}
\end{ruledtabular}
\end{table}

\begin{table}
\caption[C3v]{
 Our choices for the representation matrices $\Gamma$ of the single-valued and
 double-valued irreps $\Lambda$ of the little group $C_{3v}$ for momentum in the direction
 $(1,1,1)$.  The matrices for all other group elements can be obtained from appropriate 
 multiplications of the matrices below.  $C_{3\delta}=C_{4y}C_{4z}$ is a rotation about
$(1,1,1)$, and $C_{2b}=C_{2x}C_{4z}$ is a rotation about $(1,-1,0)$.
\label{tab:C3v_irrep_matrices}}
\begin{ruledtabular}
\begin{tabular}{ccc}
 $\Lambda$ & $ \Gamma^{\left( \Lambda \right) } \left( C_{3 \delta} 
\right) $ & $ \Gamma^{\left( \Lambda \right) } \left( I_{s} C_{2 b} \right) $ \\
\hline
$A_{1}\rule[-0mm]{0pt}{1mm}$ & $\left[ 1 \right]$ & $\left[1 \right] $ \\
$A_{2}\rule[-0mm]{0pt}{1mm}$ & $\left[ 1 \right]$ & $\left[-1 \right] $ \\
$E\rule[-3ex]{0pt}{6ex}$ & $ \displaystyle\frac{1} {2} \left[ \begin{array}{cc} -1 & 
\sqrt{3} \\ -\sqrt{3} & -1   \end{array}  \right] $  &  
$ \left[ \begin{array}{cc} -1 & 0 \\ 0 & 1 \end{array} \right] $ \\
$F_{1}\rule[-0ex]{0pt}{1ex}$ & $\left[ -1 \right]$ & $\left[ i \right] $ \\
$F_{2}\rule[-0ex]{0pt}{1ex}$ & $\left[ -1 \right]$ & $\left[ -i \right] $ \\
$G\rule[-3ex]{0pt}{7ex}$ & $ \displaystyle\frac{1} {2} \left[ \begin{array}{cc} 1-i &
 -1-i \\ 1-i & 1+i   \end{array}  \right] $  & 
 $ \displaystyle\frac{1} {\sqrt{2}} \left[ \begin{array}{cc} 0 & 1-i \\ 
-1-i & 0 \end{array} \right] $ \\
\end{tabular}
\end{ruledtabular}
\end{table}

For a given set of elemental hadron operators that transform among
one another, many of the projections in Eq.~(\ref{eq:project}) 
vanish or lead to linearly-dependent operators, so the final step in
the operator construction is to choose suitable linear combinations of
the projected operators to obtain a final set of independent single-hadron
operators.  These linear combinations are obtained using a Gram-Schmidt
procedure as described in Ref.~\cite{baryons2005A}.

At the end of this entire procedure, we obtain single-hadron annihilation 
operators $B_{\pvec\Lambda\lambda i}^{II_3S}$ characterized by total 
isospin $I$, the projection of the total isospin $I_3$, strangeness $S$, 
momentum $\pvec$, little group irrep $\Lambda$, and irrep row $\lambda$.  
Here, we use the index $i$ to indicate all other quantum numbers and 
identifying information.
All of these single-hadron operators constructed as described above
transform under a group element $(R,\bvec)$ of the space group $O_h^1$,
in which $\xvec\rightarrow R\xvec+\bvec$, according to
\begin{eqnarray}
 U_{(R,\bvec)} B_{\pvec\Lambda\lambda i}^{II_3S}(t)\, U_{(R,\bvec)}^\dagger
 \!\!\!&=& \! B_{R\pvec\,\Lambda\mu i}^{II_3S}(t)
  \, \Gamma^{(\Lambda)}_{\mu\lambda}(R_W^{\pvec})^\ast e^{i\bvec\cdot R\pvec},
 \nonumber \\
 U_{(R,\bvec)} \overline{B}_{\pvec\Lambda\lambda i}^{II_3S}(t)\, U_{(R,\bvec)}^\dagger
 \!\!\!&=&\!  \overline{B}_{R\pvec\,\Lambda\mu i}^{II_3S}(t)
 \, \Gamma^{(\Lambda)}_{\mu\lambda}(R_W^{\pvec}) e^{-i\bvec\cdot R\pvec},
  \nonumber\\
\label{eq:hadrontransform}
\end{eqnarray}
where the Wigner rotation is given by
\begin{equation}
 R_W^{\pvec} = (R_{\rm ref}^{R\pvec})^{-1}\ R \ R_{\rm ref}^{\pvec},
\label{eq:wignerrot}
\end{equation}
and is an element of the little group of $\pvec$.  Note that the above
equations apply even when $R$ refers to spatial inversion $I_s$.
Eqs.~(\ref{eq:hadrontransform}) and (\ref{eq:wignerrot})
play a crucial role when forming the multi-hadron operators.
The behaviors of our operators under $G$-parity and isospin rotations
are considered below.

In Ref.~\cite{baryons2005A}, the odd-parity zero-momentum baryons were 
constructed from their even-parity partner operators utilizing a particular
transformation involving charge-conjugation.  For a given even-parity 
zero-momentum baryon operator $B^g_i(t)$, an odd-parity operator $B^u_i(t)$ 
can be defined by rotating the three Dirac indices using the $\gamma_2$ matrix 
and replacing the expansion coefficients by their complex conjugates.
This yields particular relationships between the temporal correlation 
matrices of the even-parity baryons and the time-reversed correlation
matrices of the odd-parity baryon operators, allowing averaging over 
forward and backward temporal propagations in some cases for increased 
statistics.  However, for the baryon operators having nonzero momentum, 
parity is no longer a good quantum number since it reverses the 
three-momentum.  For this reason, we do not bother to apply a 
generalization of the above procedure in constructing the odd-parity 
baryon operators of nonzero momentum, especially since the increased 
statistical precision can be obtained in other more efficient ways when 
using the stochastic LapH method.

In constructing our light pion and kaon operators, we take symmetry under
time reversal into account.  Although the lattices we use are rather large in 
temporal extent, temporal wrap-around effects can still come into play for 
the light pions and kaons in certain situations where high precision is needed, 
such as in studying the $\pi\pi$ scattering phase shifts.  Energies can be
extracted with increased statistical precision if meson operators whose 
temporal correlations are symmetric under time reversal are used.  Symmetry 
under time reversal helps in extracting meson energies whenever temporal 
wrap-around effects become non-negligible since the functional forms used
for fitting the data have fewer parameters, leading to more precise energy
estimates.  For meson operators, the energies of the states propagating 
backwards in time are the same as those of the states traveling forwards in 
time, but the couplings of a given meson operator to the forward-propagating
states can differ from the couplings to the backward-propagating states
if symmetry under time reversal is not taken into account.

We can reduce the number of needed fit parameters if we use meson operators
whose temporal correlation matrices satisfy
$
  C_{\mu\nu}(t)=C_{\mu\nu}(N_t-t),
$
where $N_t$ is the temporal extent of the lattice, assuming periodic
boundary conditions in time.
This can be achieved if each meson operator itself satisfies
$
   M_i(t) = \eta\ M_i(N_t-t),
$
with $\vert\eta\vert^2=1$.  Under a certain symmetry of the lattice
action that involves time reversal, the covariantly-displaced
LapH-smeared quark fields transform according to
\begin{eqnarray}
   q^A_{a\alpha j}(x) &\longrightarrow&
  (\gamma_4\gamma_5)_{\alpha\beta}\ q^A_{a\beta j}({\cal T}x),
 \nonumber\\
   \overline{q}^A_{a\alpha j}(x)&\longrightarrow &
  \overline{q}^A_{a\beta j}({\cal T}x)
  \ (\gamma_4\gamma_5)_{\beta\alpha},
\label{eq:TRrestrict}
\end{eqnarray}
where $({\cal T}x)_j=x_j$ and $({\cal T}x)_4=N_t-x_4$.
Since wrap-around effects are a potential problem only for the lightest 
mesons, we decided to modify only the symmetry channels containing the
lightest pseudoscalars ($\pi, K, \eta$).  In each case, we projected the 
operators into the even and odd operators under the above transformation.  
Numerical tests using a small number of configurations showed that the odd
operators performed somewhat better, but the difference was not very
significant.  Hence, we discarded the even operators, and kept the odd 
operators.  Note that the operator $\overline{\psi}\gamma_5\psi$ is odd 
under the above transformation. 

\begin{table}[t]
\caption[captab]{Flavor structure of the elemental hadron annihiliation
 operators we use. Each is characterized by isospin $I$, maximal $I_3=I$,  
 strangeness $S$, and $G$-parity, where applicable, expressed in terms 
 of the gauge-invariant three-quark and quark-antiquark operators defined in 
 Eqs.~(\protect\ref{eq:baryonelemental}) and (\protect\ref{eq:mesonelemental}),
 respectively. $U_G$ is the quantum operator that effects a $G$-parity 
 transformation.  Recall that $f,f^\prime,h,h^\prime,b,a$ are even-parity mesons, 
 while $\eta,\eta^\prime,\omega,\phi,\rho,\pi$ are odd-parity mesons.
 Also, a flavored meson is named $K^\ast$ if its total spin $J$ and parity $P$ 
 are both odd or both even, otherwise it is named $K$.
\label{tab:flavors}}
\begin{ruledtabular}
\begin{tabular}{ccrrc}
 Hadron & $I=I_3$ & $S$ & $G$ & Annihilation operators \\ \hline
 $\Delta^{++}$ & $\frac{3}{2}$ & $0$ & & $\Phi^{uuu}_{\alpha\beta\gamma}$\\
 $\Sigma^+$ & $1$ & $-1$ && $\Phi^{uus}_{\alpha\beta\gamma}$\\
 $N^+$ &  $\frac{1}{2}$ &  $0$ &&
 $\Phi^{uud}_{\alpha\beta\gamma}-\Phi^{duu}_{\alpha\beta\gamma}$ \\
 $\Xi^0$ & $\frac{1}{2}$ & $-2$ && $\Phi^{ssu}_{\alpha\beta\gamma}$\\
 $\Lambda^0$ & $0$ & $-1$ & &
 $\Phi^{uds}_{\alpha\beta\gamma}-\Phi^{dus}_{\alpha\beta\gamma}$ \\
$\Omega^-$ & $0$ & $-3$ && $\Phi^{sss}_{\alpha\beta\gamma}$ \\
 $f,f^\prime,\eta,\eta^\prime$ & $0$ & $0$ & $1$ & 
   $\Phi^{uu}_{\alpha\beta}\!+\!\Phi^{dd}_{\alpha\beta}
   \!+\! U_G(\Phi^{uu}_{\alpha\beta}\!+\!
   \Phi^{dd}_{\alpha\beta})U_G^\dagger$\\
  & & & & $\Phi^{ss}_{\alpha\beta}+U_G
   \Phi^{ss}_{\alpha\beta}U_G^\dagger$\\
 $h,h^\prime,\omega,\phi$ & $0$& $0$ & $-1$ & 
   $\Phi^{uu}_{\alpha\beta}\!+\!\Phi^{dd}_{\alpha\beta}
   \!-\! U_G(\Phi^{uu}_{\alpha\beta}
   \!+\!\Phi^{dd}_{\alpha\beta})U_G^\dagger$\\
  & & & &$\Phi^{ss}_{\alpha\beta}-U_G
   \Phi^{ss}_{\alpha\beta}U_G^\dagger$\\
 $b^+,\rho^+$ & $1$  & $0$ &$1$ & 
   $\Phi^{du}_{\alpha\beta}+U_G
   \Phi^{du}_{\alpha\beta}U_G^\dagger$\\
 $a^+,\pi^+$ & $1$  & $0$ &$-1$ & 
   $\Phi^{du}_{\alpha\beta}-U_G
   \Phi^{du}_{\alpha\beta}U_G^\dagger$\\
 $K^+,K^{\ast +}$ & $\frac{1}{2}$ &$1$ & & $\Phi^{su}_{\alpha\beta}$\\
 $\overline{K}^0, \overline{K}^{\ast 0}$ 
& $\frac{1}{2}$ &$-1$ & & $\Phi^{ds}_{\alpha\beta}$
\end{tabular}
\end{ruledtabular}
\end{table}

As in Ref.~\cite{baryons2005A}, we work in the approximation that the 
masses of the $u$ and $d$ quarks are equal, and we neglect electromagnetic
interactions.  In this approximation, the theory has an exact isotopic spin
symmetry, and states are characterized by total isospin $I$, its projection
$I_3$ onto a given axis, and strangeness $S$.  Again, we consider only
the $u,d,s$ quarks here.  Incorporating this isospin symmetry into our
operators is straightforward and has been described in 
Ref.~\cite{baryons2005A}.  The flavor structure of the hadron operators
we use are summarized in Table~\ref{tab:flavors}.
Due to an approximate $SU(3)$ $uds$-flavor symmetry, quark flavor combinations
in meson and baryon operators are often chosen according to the irreducible
representations of $SU(3)$ flavor.  Such combinations are simply linear
superpositions of the operators presented in Table~\ref{tab:flavors}.  Since 
we plan to obtain Monte Carlo estimates of the complete correlation matrices 
of operators including all allowed flavor combinations, the use of linear
superpositions which transform irreducibly under $SU(3)$ flavor is
unnecessary.  For example, we construct isoscalar meson operators
having flavor content $\overline{u}u+\overline{d}d$ separately from those
having flavor content $\overline{s}s$.  Our correlation matrices
then allow mixings of these operators.  Our choices of operators described 
in Table~\ref{tab:flavors} are dictated by computational simplicity.
In summary, we construct single-hadron operators that transform under
an isospin rotation $R_\tau$ according to
\begin{eqnarray}  
 U_{R_\tau}\, B_{\pvec\Lambda\lambda i}^{II_3S}(t)\,U_{R_\tau}^\dagger
  &=&  B_{\pvec\Lambda\lambda i}^{II_3^\prime S}(t)
\ D^{(I)}_{I_3^\prime I_3}(R_\tau)^\ast , \nonumber \\
  U_{R_\tau}\, \overline{B}_{\pvec\Lambda\lambda i}^{II_3S}(t)\, U_{R_\tau}^\dagger
  &=& \overline{B}_{\pvec\Lambda\lambda i}^{II_3^\prime S}(t)
\ D^{(I)}_{I_3^\prime I_3}(R_\tau),
\label{eq:isospindef}
\end{eqnarray}
where $D^{(I)}(R_\tau)$ are the familiar Wigner rotation matrices.
Our meson operators $M$ are constructed such that they transform 
under $G$-parity according to
\begin{eqnarray}
 U_G\, M_{\pvec\Lambda\lambda i}^{II_3,S}(t)\, U_G^\dagger
 &=& \eta_\Lambda M_{\pvec\,\Lambda\lambda i}^{II_3,-S}(t),
 \nonumber \\
 U_G\, \overline{M}_{\pvec\Lambda\lambda i}^{II_3,S}(t)\, U_G^\dagger
 &=&  \eta_\Lambda\overline{M}_{\pvec\,\Lambda\lambda i}^{II_3,-S}(t),
\label{eq:mesontransformG}
\end{eqnarray}
where $\eta_\Lambda=1$ if $S=\pm 1$, and when the strangeness $S=0$, then
$\eta_\Lambda=\pm 1$ depending
on the $G$-parity superscript of the irrep $\Lambda$.

\begin{table}
\caption[jspinboson]{Continuum limit spin identification:
   the number $n_\Lambda^J$ of times that the $\Lambda$ single-valued irrep 
   of the octahedral point group $O_h$ occurs in the
   (reducible) subduction of the integer $J$ irrep of $SU(2)$.  The numbers
   for $A_{1u}$, $A_{2u}$, $E_u$, $T_{1u}$, $T_{2u}$ are the same as for 
   $A_{1g}$, $A_{2g}$, $E_g$, $T_{1g}$, $T_{2g}$, respectively.
\label{tab:bosonspin}}
\begin{ruledtabular}
\begin{tabular}{lrrrrr}
   $J$ & $n^J_{A_{1g}}$ & $n^J_{A_{2g}}$ & $n^J_{E_g}$ & $n^J_{T_{1g}}$ 
   & $n^J_{T_{2g}}$\\ \hline
  0  & $ 1$ & $ 0$ & $ 0$ & $ 0$ & $ 0$ \\
  1  & $ 0$ & $ 0$ & $ 0$ & $ 1$ & $ 0$ \\
  2  & $ 0$ & $ 0$ & $ 1$ & $ 0$ & $ 1$ \\
  3  & $ 0$ & $ 1$ & $ 0$ & $ 1$ & $ 1$ \\
  4  & $ 1$ & $ 0$ & $ 1$ & $ 1$ & $ 1$ \\
  5  & $ 0$ & $ 0$ & $ 1$ & $ 2$ & $ 1$ \\
  6  & $ 1$ & $ 1$ & $ 1$ & $ 1$ & $ 2$ \\
  7  & $ 0$ & $ 1$ & $ 1$ & $ 2$ & $ 2$ \\
  8  & $ 1$ & $ 0$ & $ 2$ & $ 2$ & $ 2$ \\
  9  & $ 1$ & $ 1$ & $ 1$ & $ 3$ & $ 2$
\end{tabular}
\end{ruledtabular}
\end{table}

\begin{table}
\caption[jspinfermion]{Continuum limit spin identification:
   the number $n_\Lambda^J$ of times that the $\Lambda$ double-valued irrep
   of the octahedral point group $O_h$ occurs in the (reducible) subduction 
   of the half-integral $J$ irrep of $SU(2)$.  The numbers
   for $G_{1u},G_{2u},H_u$ are the same as for $G_{1g},G_{2g},H_g$,
   respectively.
\label{tab:fermionspin}}
\begin{ruledtabular}
\begin{tabular}{cccc@{\hspace{3em}}cccc}
  $J$  & $n^J_{G_{1g}}$ & $n^J_{G_{2g}}$ & $n^J_{H_g}$ &
 $J$  & $n^J_{G_{1g}}$ & $n^J_{G_{2g}}$ & $n^J_{H_g}$\\ \hline
  $\frac{1}{2}$  &  $1$ & $0$ & $0$ &$\frac{9}{2} $ &  $1$ & $0$ & $2$ \\
  $\frac{3}{2}$  &  $0$ & $0$ & $1$ &$\frac{11}{2}$ &  $1$ & $1$ & $2$ \\
  $\frac{5}{2} $ &  $0$ & $1$ & $1$ &$\frac{13}{2}$ &  1 & 2 & 2 \\
  $\frac{7}{2} $ &  $1$ & $1$ & $1$ &$\frac{15}{2}$ &  1 & 1 & 3
\end{tabular}
\end{ruledtabular}
\end{table}

To associate our finite-box energies with observed hadrons,
it is necessary to know which spin-$J$ irreps of the continuous group
of rotations occur in which irreps of the octahedral point group.  
For the convenience of the reader, the spin contents of the $O_h$ 
single-valued and double-valued irreps are listed in 
Tables~\ref{tab:bosonspin} and \ref{tab:fermionspin}, respectively. 
These tables list the number of times that each irrep $\Lambda$ of $O_h$ 
appears in various $J$ irreps of $SU(2)$ subduced to the double group
of $O_h$.  Table~\ref{tab:movingsubduction} is useful for identifying
which hadrons of nonzero momentum appear in which irreps of the little 
groups. The decompositions of the subduced representations of $O_{h}$ 
into the irreps of the little groups $C_{4v}$, $C_{3v}$, and $C_{2v}$
are given in this table.  Table~\ref{tab:particles} lists the irreps 
of $O_h$ in which various common hadrons at rest appear.

\begin{table}
\caption{Subduction $\downarrow$ of the irreducible representations of 
$O_{h}$ to the irreducible representations of the little groups $C_{4v}$, 
$C_{3v}$, and $C_{2v}$.}
\label{tab:movingsubduction}
\begin{ruledtabular}
\begin{tabular}{cccc}
$\Lambda\left(O_{h} \right)$ & $  \downarrow C_{4 v}$ 
    & $  \downarrow C_{3 v}$
    & $  \downarrow C_{2 v}$\\
\hline
$A_{1 g}$    &  $A_{1}$              &  $A_{1}$                        &  $A_{1}$                            \\
$A_{1 u}$    &  $A_{2}$              &  $A_{2}$                        &  $A_{2}$                            \\
$A_{2 g}$    &  $B_{1}$              &  $A_{2}$                        &  $B_{2}$                            \\
$A_{2 u}$    &  $B_{2}$              &  $A_{1}$                        &  $B_{1}$                            \\
$E_{g}$      &  $A_{1} \oplus B_{1}$ &  $E$                            &  $A_{1} \oplus B_{2}$               \\
$E_{u}$      &  $A_{2} \oplus B_{2}$ &  $E$                            &  $A_{2} \oplus B_{1}$               \\
$T_{1 g}$    &  $A_{2} \oplus E$     &  $A_{2} \oplus E$               &  $A_{2} \oplus B_{1} \oplus B_{2}$  \\
$T_{1 u}$    &  $A_{1} \oplus E$     &  $A_{1} \oplus E$               &  $A_{1} \oplus B_{1} \oplus B_{2}$  \\
$T_{2 g}$    &  $B_{2} \oplus E$     &  $A_{1} \oplus E$               &  $A_{1} \oplus A_{2} \oplus B_{1}$  \\
$T_{2 u}$    &  $B_{1} \oplus E$     &  $A_{2} \oplus E$               &  $A_{1} \oplus A_{2} \oplus B_{2}$  \\
$G_{1 g/u}$  &  $G_{1}$              &  $G$                            &  $G$                                \\
$G_{2 g/u}$  &  $G_{2}$              &  $G$                            &  $G$                                \\
$H_{g/u}$    &  $G_{1} \oplus G_{2}$ &  $F_{1} \oplus F_{2} \oplus G$  &  $2G$  
\end{tabular}
\end{ruledtabular}
\end{table}

\begin{table}
\caption[particles]{The irreducible representations of $O_h$
in which various commonly-known hadrons at rest occur.
\label{tab:particles}}
\begin{ruledtabular}
\begin{tabular}{lc@{\hspace{3em}}lc@{\hspace{3em}}lc}
Hadron & Irrep & Hadron & Irrep & Hadron & Irrep \\ \hline
$\pi$ & $A_{1u}^-$ &  $K$ & $A_{1u}$ & $\eta,\eta^\prime$ & $A_{1u}^+$\\
$\rho$ & $T_{1u}^+$ & $\omega,\phi$ & $T_{1u}^-$ & $K^\ast$ & $T_{1u}$\\
$a_0$ & $A_{1g}^+$ & $f_0$ & $A_{1g}^+$ & $h_1$ & $T_{1g}^-$\\
$b_1$ & $T_{1g}^+$ & $K_1$ & $T_{1g}$ & $\pi_1$ & $T_{1u}^-$\\
$N,\Sigma$ & $G_{1g}$ & $\Lambda,\Xi$ & $G_{1g}$ & $\Delta,\Omega$ & $H_g$
\end{tabular}
\end{ruledtabular}
\end{table}

\section{Implementation details}
\label{sec:implement}

In order to test the effectiveness of the single-hadron operators
that we have designed, Monte Carlo calculations must be carried out.
Details on how the temporal correlations of hadron operators are
evaluated using Monte Carlo integration with the stochastic LapH method 
are presented in this section.  Many of these details have already been 
described in Ref.~\cite{Morningstar:2011ka}, so only details in addition 
to those in Ref.~\cite{Morningstar:2011ka} are presented here, along with
some reiteration of important run parameters.

Our computations use the so-called stochastic LapH method\cite{Morningstar:2011ka}
and are done in a sequence of steps: (a) generation of
gauge-field configurations using the Monte Carlo method; (b) computation
of quark sinks for various noises and dilution projectors using the
configurations from the first step; (c) computation of the single meson and 
single baryon sources and sinks using the quark sources and sinks from the 
second step; (d) evaluation of the correlators using the single-hadron sources 
and sinks and their combinations into multi-hadron sources and sinks; and 
(e) analysis of the correlators to extract the energies.  These steps
are discussed below.

A description of step (a) is given in Ref.~\cite{Lin:2008pr}.
We are currently focusing on three Monte Carlo ensembles: (A) a set of 
412 gauge-field configurations on a large $32^3\times 256$ anisotropic lattice 
with a pion mass $m_\pi\sim 240$~MeV, (B) an ensemble of 551 configurations
on an $24^3\times 128$ anisotropic lattice with a pion mass
$m_\pi\sim 390$~MeV, and (C) an ensemble of 584 configurations
on an $24^3\times 128$ anisotropic lattice with a pion mass
$m_\pi\sim 240$~MeV.  We refer to these ensembles as the 
$(32^3\vert 240)$, $(24^3\vert 390)$, and $(24^3\vert 240)$ ensembles,
respectively.  These ensembles were generated using the Rational 
Hybrid Monte Carlo (RHMC) algorithm\cite{Clark:2004cp}, which is a 
Metropolis method with a sophisticated means of proposing a global change 
to the gauge and pseudofermion fields. A fictitious momentum is introduced
for each link variable with a Gaussian distribution, and a Hamiltonian
is formed involving these momenta and the original action as a potential 
energy. A new field configuration is proposed by approximately solving 
Hamilton's equations for some length of fictitious time, known as an RHMC 
trajectory. In each ensemble, successive 
configurations are separated by 20 RHMC trajectories to minimize autocorrelations.
An improved anisotropic clover fermion action and an improved gauge field 
action are used\cite{Lin:2008pr}.  In these ensembles, $\beta=1.5$
and the $s$ quark mass parameter is set to $m_s=-0.0743$ in order to reproduce 
a specific combination of hadron masses\cite{Lin:2008pr}.  
In the $(24^3\vert 390)$ ensemble, the light quark mass parameters are set to
$m_u=m_d=-0.0840$ so that the pion mass is around 390~MeV if one sets the scale 
using the $\Omega$ baryon mass.   In the $(32^3\vert 240)$ and 
$(24^3\vert 240)$ ensembles, $m_u=m_d=-0.0860$ are used, resulting in a pion 
mass around 240~MeV.  The spatial grid size is $a_s\sim 0.12$~fm, whereas
the temporal spacing is $a_t\sim 0.035$~fm.

A description of step (b) above, computation of the quark sinks, is given 
in Ref.~\cite{Morningstar:2011ka}.  We employ the Laplacian Heaviside (LapH) 
quark-field smearing scheme defined using the three-dimensional 
gauge-covariant Laplacian expressed in terms of a stout-smeared gauge field.
The spatial links are smeared using the stout-link procedure described in 
Ref.~\cite{Morningstar:2003gk} with $n_\xi=10$ iterations and staple weight 
$\xi=0.10$.  For the cutoff in the LapH smearing, we use $\sigma_s^2=0.33$, 
which translates into the number $N_v$ of LapH eigenvectors retained
being $N_v=112$ for the $24^3$ lattices and $N_v=264$ for the $32^3$ lattice.
We use $Z_4$ noise in all of our stochastic estimates of quark propagation.
Our variance reduction procedure is similar to that described in Ref.~\cite{Foley:2005ac}.
Our noise dilution projectors are products of time dilution, spin dilution, 
and LapH eigenvector dilution projectors.  We use a triplet
(T, S, L) to specify a given dilution scheme, where ``T" denotes time,
``S" denotes spin, and ``L" denotes LapH eigenvector dilution.  The schemes
are denoted by 1 for no dilution, F for full dilution, and B$K$ and I$K$ for
block-$K$ and interlace-$K$, respectively (see Ref.~\cite{Morningstar:2011ka}).  
For all forward-time quark lines connecting source time $t_0$ to the later
sink time $t_F$, we use the dilution scheme (TF, SF, LI$j$), where $j=8$ for mesons
and $j=4$ for baryons.  We chose $j=4$ for baryons in order to dramatically
reduce the disk space needed to store the baryon sources/sinks by a factor of eight.
We found that the statistical errors in the baryon effective masses only increased by 
a factor of two in changing from LI8 to LI4.  Given the eightfold reduction in 
both storage and computing time, we deemed this an acceptable loss of accuracy.
For all same-sink-time $t_F$-to-$t_F$ quark lines, the dilution scheme (TI16, SF, LI$j$) 
is used, where $j=8$ for mesons and $j=4$ for baryons. The TI16 interlacing in time 
enables us to evaluate quark lines that originate on \textit{any} time slice, allowing 
us to evaluate all diagrams needed to obtain the temporal correlations involving
single-hadron and multi-hadron operators.   Four widely-separated source times 
$t_0$ are used on each $24^3$ gauge configuration, whereas eight $t_0$ values 
are used on the $32^3$ lattice.  

Details about how the single-hadron sources and sinks are computed in
step (c) above can be found in Ref.~\cite{Morningstar:2011ka}.
A comprehensive survey of the spectrum of excited states in QCD
requires obtaining temporal correlations of a large number of
different operators.  For example, $K\overline{K}$ operators must
mix with $\pi\pi$ operators, and so on.  Hence, a large variety
of single-hadron operators must be available in the study of each
sector, and a plethora of Wick contractions must be evaluated.  
In order to study all stationary states
of QCD involving the $u,d,s$ quarks, we need suitable sources and sinks
for all isovector mesons ($\pi,\rho,a,b$), all isoscalar mesons 
($\eta,f,h,\omega,\phi)$, and all kaons.  To study $\overline{K}K$ states, 
we also need separate antikaon sources and sinks.   For the isoscalar mesons, 
separate $\overline{u}u+\overline{d}d$ and $\overline{s}s$ operators are 
evaluated, and these will be allowed to mix in the correlation matrices.
We also need all single-baryon $N,\Delta,\Xi,\Lambda,\Sigma,\Omega$ sources 
and sinks.  Each hadron source and sink involves a summation over color 
indices and the spatial sites on each time-slice of the lattice.
Different spin components and displacement directions are combined
to form the hadron operators which transform irreducibly under the
symmetry operations of the spatial lattice.

\begin{table}
\caption{The numbers $N_{\rm op}$ of single-hadron operators whose sinks and sources 
have been computed.  The operator numbers include operators having zero, on-axis, 
planar-diagonal, and cubic-diagonal momenta, as well as momenta for the pions in 
various special directions, such as $(0,1,2)$ and $(1,1,2)$.  
$N_{\rm mom}$ indicates the number of momenta in the operator
sets. Here, $\pi$ refers to \textit{any} isovector quark-antiquark meson 
operator with flavor content such as $\overline{d}u$ (such as $a$, $b$, $\pi$, and 
$\rho$ mesons), $\eta$ refers to any $\overline{u}u+\overline{d}d$ isoscalar meson 
operator, $\phi$ refers to any $\overline{s}s$ isoscalar meson operator,  $K$ refers 
to any quark-antiquark operator having flavor content $\overline{s}u$ or 
$\overline{s}d$ such that its strangeness is $S=1$, and $\overline{K}$ is any 
quark-antiquark operator having flavor content $\overline{u}s$ or $\overline{d}s$ 
such that its strangeness is $S=-1$.   
%Four widely-separated source times $t_0$ are used on the 551 configurations of 
%the $(24^3\vert 390)$ ensemble and the 584 configurations of the $(24^3\vert 240)$ 
%ensemble, whereas eight $t_0$ values are used on the 412 configurations of the 
%$(32^3\vert 240)$ ensemble.
\label{tab:operators}}
\begin{ruledtabular}
\begin{tabular}{crlrlrl}
 & \multicolumn{2}{c}{$(24^3\vert 390)$} 
  &\multicolumn{2}{c}{$(24^3\vert 240)$} &
  \multicolumn{2}{c}{$(32^3\vert 240)$} \\
  & $N_{\rm op}$ & $N_{\rm mom}$ & $N_{\rm op}$ & $N_{\rm mom}$ 
  & $N_{\rm op}$ & $N_{\rm mom}$ \\ \hline
 $\pi$           & 1776 &123 & 2028 &123 & 2740 &149 \\
 $\eta$          & 2012 &51 & 2204 & 51 & 3078 & 77 \\
 $\phi$          & 2012 &51 & 2204 & 51 & 3078 & 77 \\
 $K$             & 1499 &51 & 1517 & 51 & 1949 & 65 \\
 $\overline{K}$  & 1499 &51 & 1517 & 51 & 1949 & 65 \\
 $N/\Delta$      & 1472 &33 & - & - & 1616 & 59 \\
 $\Lambda/\Sigma$& 2274 &33 & - & - & 2054 & 51 \\
 $\Xi$           & 1320 &33 & - & - & 700  & 51 \\
 $\Omega$        &  728 &33 & - & - & 680  & 45 
\end{tabular}
\end{ruledtabular}
\end{table}

Using the lowest-lying energies in each symmetry sector as determined 
from low-statistics runs on small $16^3$ lattices, we systematically 
identified all momenta in the different little group irreps that would 
be needed to capture the energy spectrum up to $0.5a_t^{-1}$, where $a_t$ 
is the temporal lattice spacing.  This energy value ensures that a sufficient
number of hadrons accessible to experiments can be studied without
overwhelming the computational resources available to us.
The numbers $N_{\rm op}$ of single-hadron operators whose sinks and sources 
have been computed are listed in Table~\ref{tab:operators}.   The operator 
numbers include operators having zero, on-axis, planar-diagonal, and 
cubic-diagonal momenta, as well as momenta for pions in various special directions, such 
as $(0,1,2)$ and $(1,1,2)$.  $N_{\rm mom}$ indicates the number of momenta in 
the operator sets.  These operators were chosen from much larger sets.  
Small-volume, low-statistics runs were done to select operators having
lower statistical noise and smaller contamination from higher-lying 
eigenstates.  In this table, $\pi$ refers to \textit{any} isovector quark-antiquark 
meson operator with flavor content such as $\overline{d}u$ (such as $a$, $b$, 
$\pi$, and $\rho$ mesons), $\eta$ refers to any $\overline{u}u+\overline{d}d$ 
isoscalar meson operator, $\phi$ refers to any $\overline{s}s$ isoscalar meson 
operator,  $K$ refers to any quark-antiquark operator having flavor content 
$\overline{s}u$ or $\overline{s}d$ such that its strangeness is $S=1$, and 
$\overline{K}$ is any quark-antiquark operator having flavor content 
$\overline{u}s$ or $\overline{d}s$ such that its strangeness is $S=-1$.  
The $N$ and $\Delta$ were computed simultaneously, and so were the $\Lambda$ and 
$\Sigma$, since these baryons share many of the same three-quark components.  

Our final correlation matrices will not use such large numbers of 
operators.  We will combine the single-hadron operators into two- and 
three-particle operators, projected into the irreps of the $O_h$ point group.  
A last round of operator selections will then occur.  Many of the multi-hadron 
operators will not produce signals of suitable quality or be linearly dependent with other 
operators.  We will have to identify a final set of multi-hadron operators that 
best allows us to extract the low-lying energy spectrum in each symmetry sector.

For baryons containing identical quark flavors, the sources/sinks defined by
Eq.~(23) in Ref.~\cite{Morningstar:2011ka} can be evaluated by assigning the quark 
spin and displacement indices to the identical-flavor quark lines using either a 
single canonical one-to-one mapping or by averaging over all such one-to-one mappings.  
Averaging over the quark-line mappings is tantamount to averaging over different 
noises, so a reduction in the correlator variances can occur.  However,
operators of this form have more three-quark terms in them, and the resulting 
increase in the number of ``elemental" three-quark operators that must be evaluated
raises the computational costs.  We did some small lattice studies to compare the 
benefits/cost of the two different ways of assigning spin and displacement indices
to the identical-flavor quark lines.  We decided that the modest increase in 
computational cost was warranted, given the variance reduction that resulted.

Since our hadrons are made out of spatially-displaced quark fields, we
carried out a study of the role of the displacement length in 
the meson and baryon operators.  Excited-state contamination in the
correlators of various meson and baryon operators was investigated
for a range of displacement lengths.  For the value of $a_s\sim 0.12$~fm in the
ensembles used, a displacement length of $3a_s$ produced operators having 
somewhat better overlaps with the low-lying mesons of interest, whereas 
displacement lengths $2a_s$ and $3a_s$ worked well for baryons.  Our
conclusions concerning these lengths were found to be insensitive to 
changes in the quark-field smearing, although we did not study this in
too much detail.  These displacements can be rather costly for baryons, so we 
decided to use $2a_s$ for all baryon operators.  In addition to achieving overlaps
with the states of interest, another concern in setting the displacement length 
is to obtain operators that are sufficiently different from one another to 
produce correlation matrices with reasonable condition numbers.

An outline of how the temporal correlations are obtained from the hadron 
sources and sinks, step (d) above, is given in Ref.~\cite{Morningstar:2011ka}.
First, thousands of single-hadron operators having different momenta for 
different noises, dilution schemes, and types of quark line ends 
(source or sink, normal mode or $\gamma_5$ Hermitian conjugate mode
using Eq.~(17) in Ref.~\cite{Morningstar:2011ka}) are evaluated and stored.
The second task in step (d) is computing the temporal correlations using
expressions such as Eqs.~(24) and (33) in Ref.~\cite{Morningstar:2011ka}.
Given the large number of correlators that we will need to evaluate, it is 
important to automate the Wick contraction process as much as possible.  
Maple is used to create the software for the numerical evaluations.
The first step in the Maple calculations is forming appropriate flavor combinations 
of symbolic Grassmann variables representing single-hadron and multi-hadron 
operators of definite total isospin $I$, strangeness $S$, and isospin projection 
$I_3$, raising and lowering the isospin projections of the individual constituent 
hadrons as needed.  Next, the Maple code carries out the actual Wick contractions 
in terms of the Grassmann symbols.  Finally, our Maple program outputs the C++ 
subroutines for use in numerically evaluating the correlators in terms of the 
stored hadron sources and sinks in the different flavor sectors.  In evaluating the 
individual quark-line diagrams, we apply $\gamma_5$ 
Hermiticity (see Eq.~(16) in Ref.~\cite{Morningstar:2011ka}) in cases where a 
$\psi(t_0)$ at the source connects with a $\overline{\psi}(t_F)$ at the sink. 
For all same-time quark lines, we average over estimates obtained using both 
Eqs.~(15) and (16) in Ref.~\cite{Morningstar:2011ka} to increase statistics.

\begin{figure*}
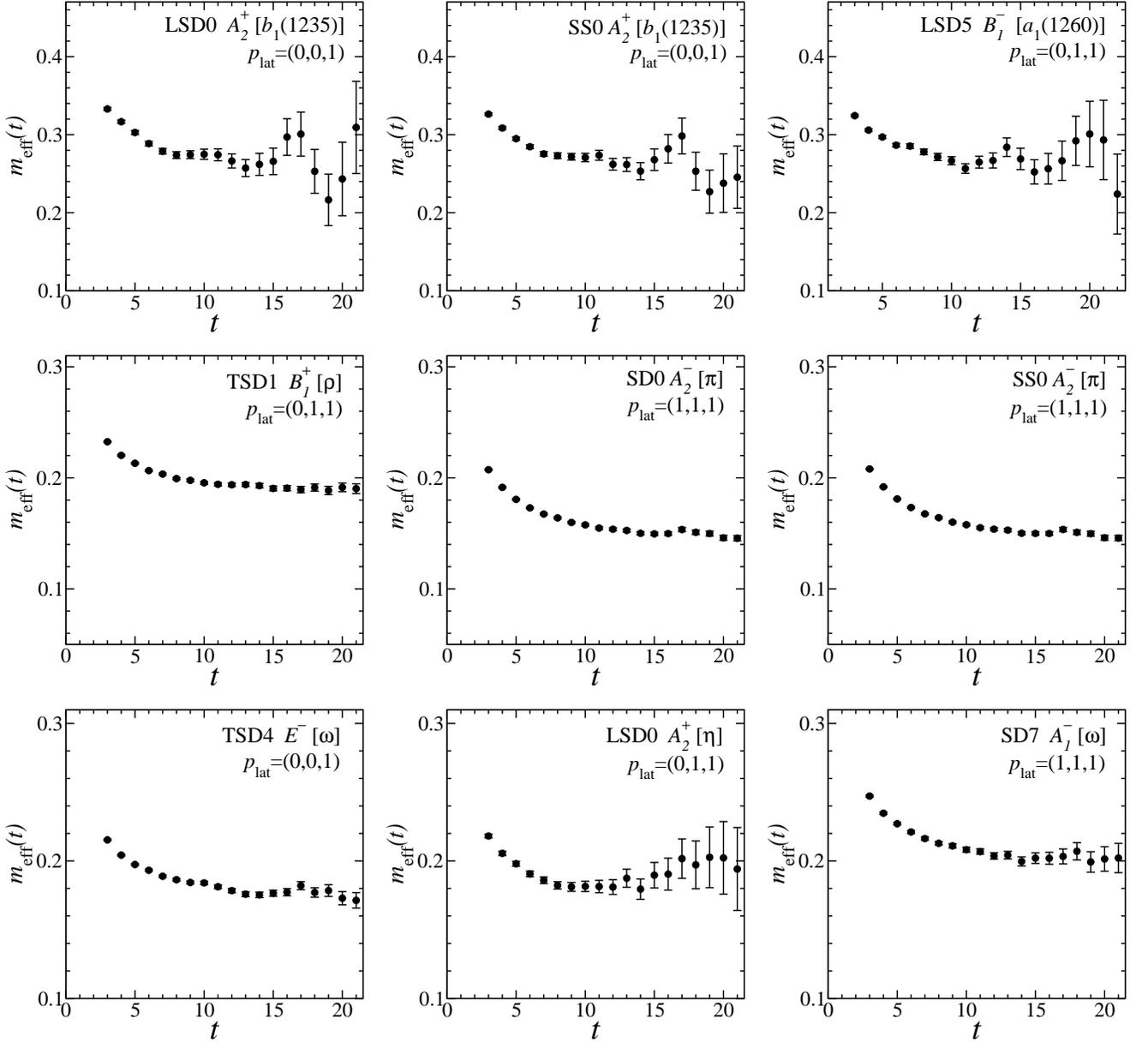

\begin{center}
\begin{minipage}{7.0in}
\includegraphics[width=2.2in,bb=7 30 541 523]{isovector1.eps}\quad
\includegraphics[width=2.2in,bb=7 30 541 523]{isovector2.eps}\quad
\includegraphics[width=2.2in,bb=7 30 541 523]{isovector3.eps}\\[3mm]
\includegraphics[width=2.2in,bb=7 30 541 523]{isovector4.eps}\quad
\includegraphics[width=2.2in,bb=7 30 541 523]{isovector5.eps}\quad
\includegraphics[width=2.2in,bb=7 30 541 523]{isovector6.eps}\\[3mm]
\includegraphics[width=2.2in,bb=7 50 541 523]{isoscalar1.eps}\quad
\includegraphics[width=2.2in,bb=7 50 541 523]{isoscalar2.eps}\quad
\includegraphics[width=2.2in,bb=7 50 541 523]{isoscalar3.eps}
\end{minipage}
\end{center}
\caption{
Effective masses, $m_{\rm eff}(t)$ using $dt=3$, associated with 
several isovector and isoscalar meson operators of various momenta 
$\bm{p}=(2\pi/L)\ \bm{p}_{\rm lat}$ on the $(24^3\vert 390)$ ensemble,
where $L=24a_s$ is the spatial extent of the lattice.  
SS refers to a single-site meson operator, and LSD denotes an operator
in which the quark is displaced from the antiquark in a longitudinal direction
along the direction of the momentum, as explained in Sec.~\ref{sec:singlehadron}.  
In a TSD operator, the quark is displaced in a direction transverse to that of 
the momentum.  For cubic-diagonal momenta, SD denotes a singly-displaced operator.
The numbers following the letters, such as in SS0, are simply 
identifying integer indices.  Particle names in the square brackets have
been included to lend context to the little group irrep labels.
\label{fig:movingmesons}}
\end{figure*}

\section{Tests of the single-hadron operators}
\label{sec:singletests}

\begin{figure*}
\begin{center}
\begin{minipage}{7.0in}
\includegraphics[width=2.1in,bb=1 10 477 477]{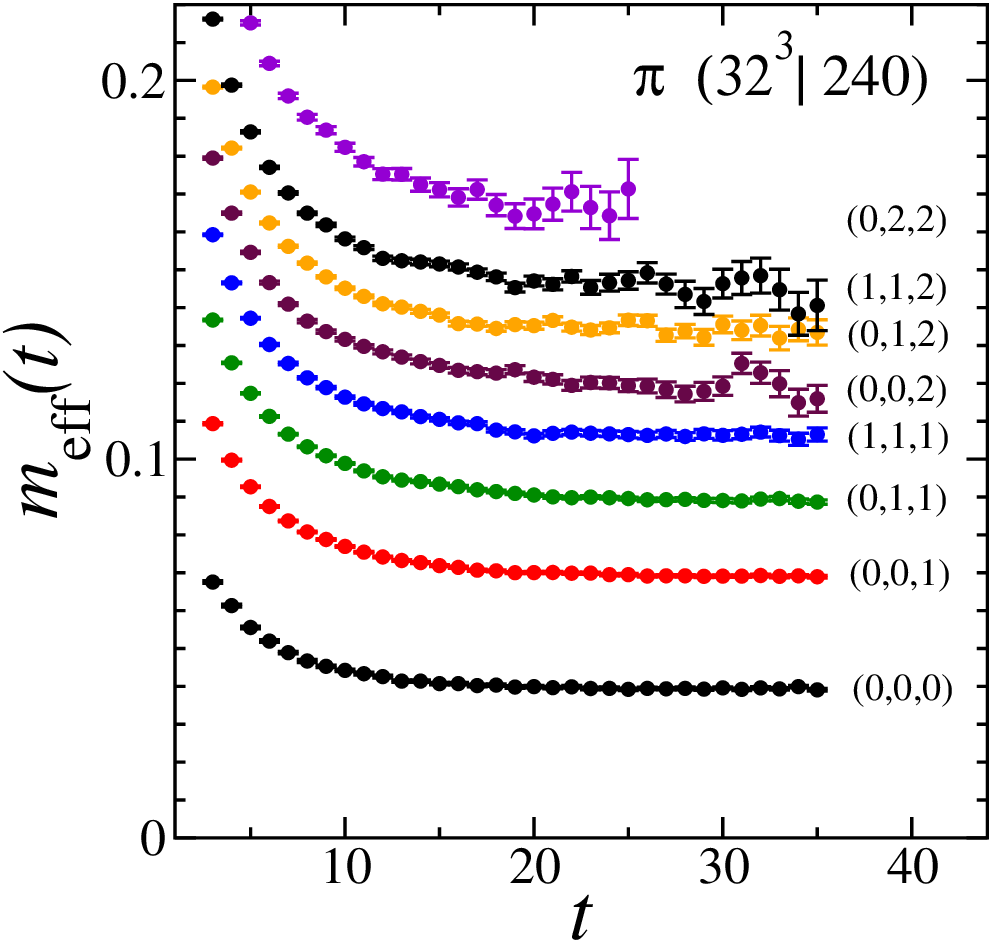}\quad
\includegraphics[width=2.3in,bb=8 38 572 541]{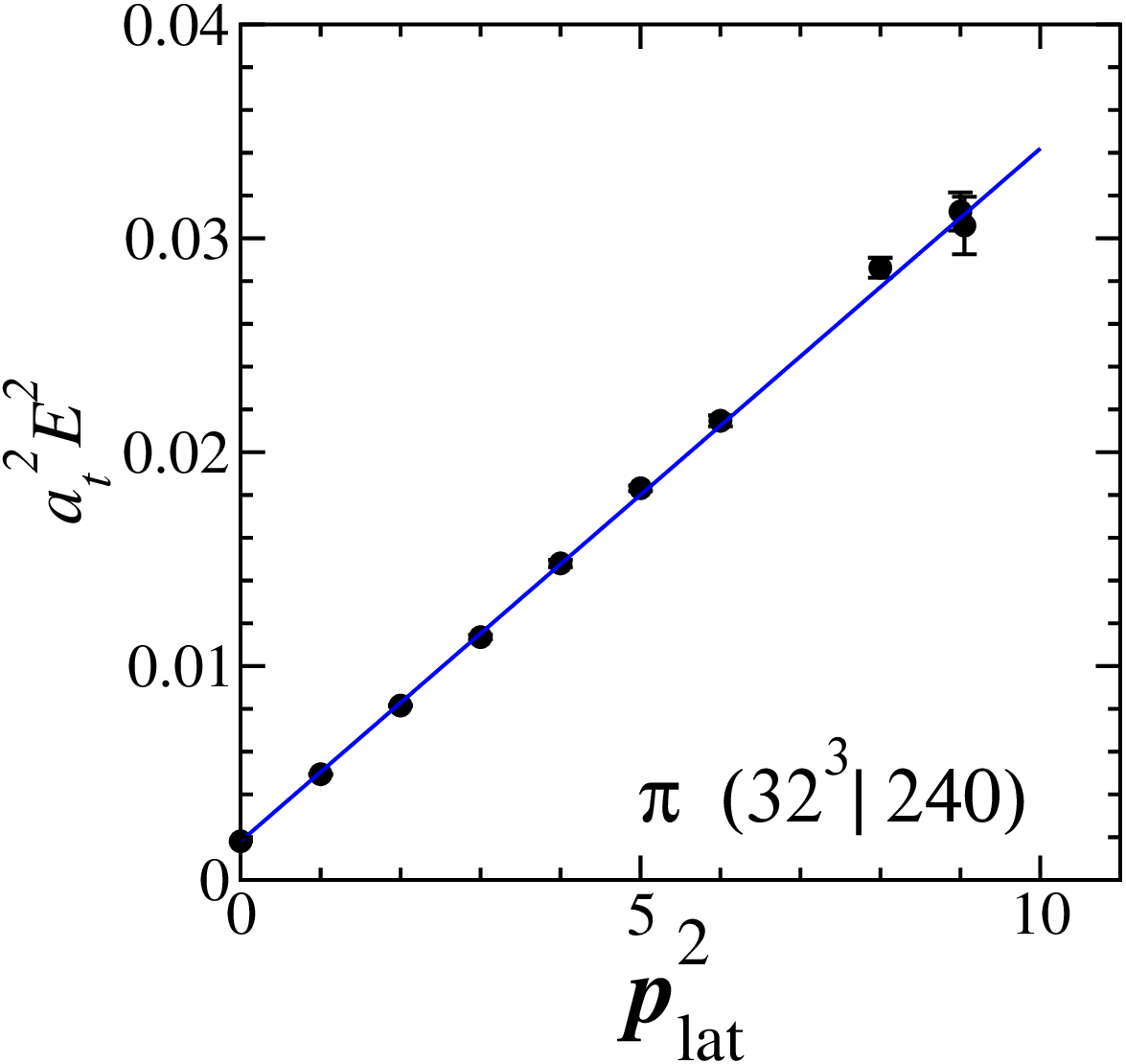}\quad
\includegraphics[width=2.25in,bb=20 20 565 523]{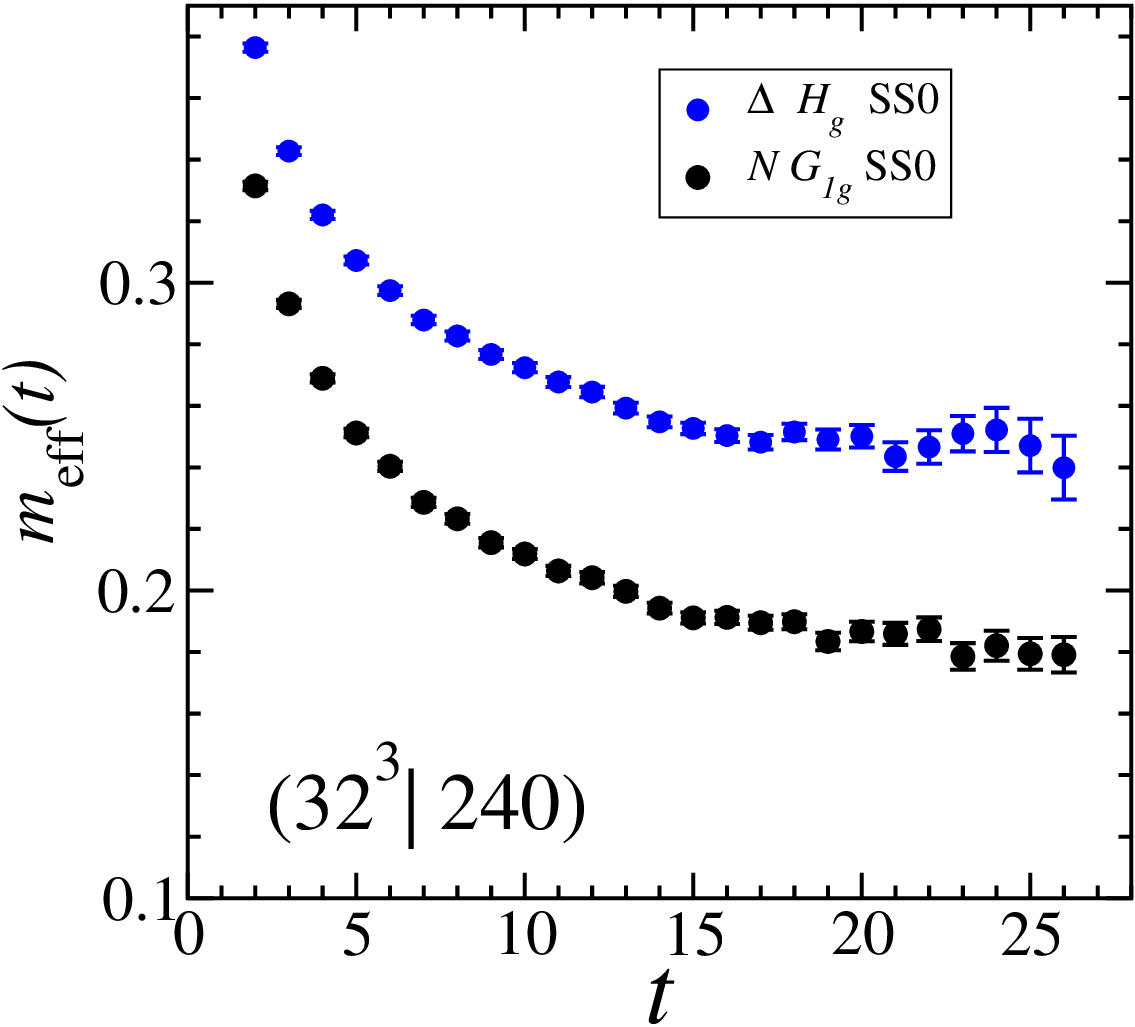}
\end{minipage}
\end{center}
\caption{
Pion and baryon results obtained using the $(32^3\vert 240)$ ensemble.
(Left) Effective masses, $m_{\rm eff}(t)$ using $dt=3$, against temporal 
separation $t$, associated with single-site pion operators having different 
momenta.  Each effective mass is labelled by its $\bm{p}_{\rm lat}$. 
(Center) Dispersion relation for the pion showing $a_t^2E^2$ against 
$\bm{p}_{\rm lat}^2$, where $E$ is the pion energy.
(Right) Effective masses, $m_{\rm eff}(t)$ 
using $dt=3$, against temporal separation $t$, associated with two single-site 
baryon operators having zero momentum.  One effective mass corresponds to the 
nucleon ($G_{1g}$ channel) and the other to a $\Delta$ baryon ($H_g$ channel).
\label{fig:piondispersion}}
\end{figure*}

In order to test the effectiveness of the single-hadron operators
that we have designed, we examined the effective masses associated with
the correlators of a variety of meson and baryon operators having various
momenta.  Samples of these effective masses are shown in 
Figs.~\ref{fig:movingmesons} and \ref{fig:piondispersion}.  In these
figures, we use the following definition of the effective mass:
\begin{equation}
  m_{\rm eff}(t) = -\frac{1}{dt}\ln\left(\frac{C(t+dt)}{C(t)}\right),
\end{equation}
where time separation $t$ in the correlator $C(t)$ is measured in term of the 
temporal lattice spacing $a_t$, and usually $dt=3$ is used.

Fig.~\ref{fig:movingmesons} shows the effective masses, $m_{\rm eff}(t)$ using 
$dt=3$, associated with several isovector and isoscalar meson operators of various 
momenta $\bm{p}=(2\pi/L)\ \bm{p}_{\rm lat}$ on the $(24^3\vert 390)$ ensemble,
where $L=24a_s$ is the spatial extent of the lattice.  
SS refers to a single-site meson operator, and LSD denotes an operator
in which the quark is displaced from the antiquark in a longitudinal direction
along the direction of the momentum, as explained in Sec.~\ref{sec:singlehadron}.  
In a TSD operator, the quark is displaced in a direction transverse to that of 
the momentum.  Several different operators of each spatial type can be
constructed, and we label these different operators using an integer that
varies from zero to the number of such operators less one.  This integer
is placed at the end of the spatial type label, such as SS0 or TSD4.
Results for on-axis momentum $\bm{p}_{\rm lat}=(0,0,1)$,
planar-diagonal momentum $\bm{p}_{\rm lat}=(0,1,1)$, and cubic-diagonal
momentum $\bm{p}_{\rm lat}=(1,1,1)$ are shown.  The names of the lowest-lying 
known particles which appear in these finite-volume symmetry channels are 
indicated in square brackets.

Energies of the pion, nucleon, and $\Delta$, obtained using the 
$(32^3\vert 240)$ ensemble, are studied in Fig.~\ref{fig:piondispersion}.  In the 
left plot, effective masses associated with single-site pion 
operators having different momenta are shown against temporal separation $t$ 
using $dt=3$.  Each effective mass is labelled by its $\bm{p}_{\rm lat}$. 
The dispersion relation for the pion is shown in the center plot, which
displays $a_t^2E^2$ against $\bm{p}_{\rm lat}^2$, where $E$ is the pion energy. 
Effective masses associated with two single-site baryon operators having zero 
momentum are shown in the right plot of Fig.~\ref{fig:piondispersion}.  One 
effective mass corresponds to the nucleon ($G_{1g}$ channel) and the other 
to a $\Delta$ baryon ($H_g$ channel).

Both Figs.~\ref{fig:movingmesons} and \ref{fig:piondispersion}
show that the stochastic LapH method works well for moving hadrons
on large lattices.  These plots also confirm that our choices of the
gauge-field and quark-field smearing parameters and our noise dilution schemes
for reducing variances are suitable for treating hadron states having
nonzero definite momenta on the three Monte Carlo ensembles we plan to use.

\section{Two-hadron operators of definite momentum}
\label{sec:twohadron}

We construct our two-hadron operators using the same procedure described
in Sec.~\ref{sec:singlehadron} to build the single-hadron operators, except
that the basic building blocks are now the single-hadron operators
instead of the elemental three-quark and quark-antiquark operators.
This procedure allows us to very efficiently build up the many multi-hadron 
operators that our spectrum computations will need.  Evaluating the 
single-hadron sources and sinks of various momenta requires summations
over both quark color and spin indices, as well as summations over the
spatial sites of the lattice.  Once these relatively expensive 
computations are done for the single hadrons, the multi-hadron operators,
being simple linear combinations of the single-hadron operators, are very
inexpensive to compute, and many of them can be quickly made.  

\begin{figure*}
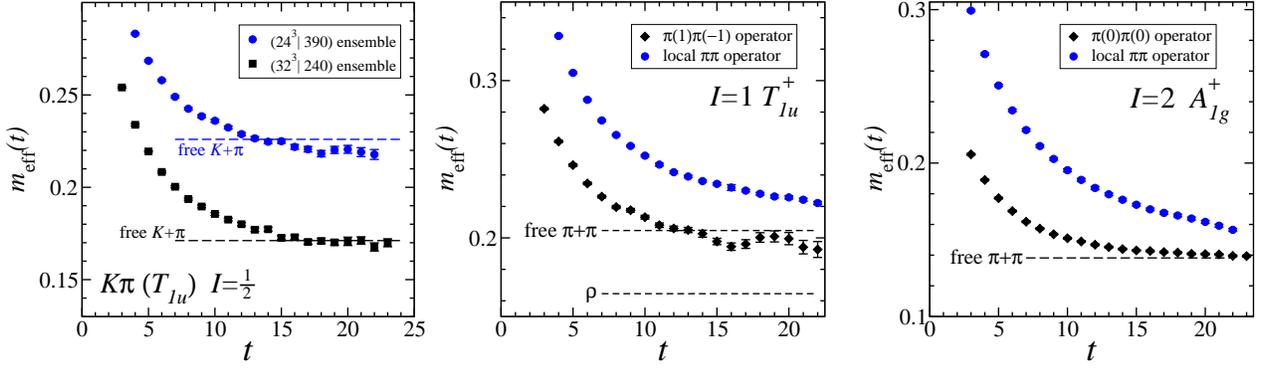

\begin{center}
\includegraphics[width=2.2in,bb=6 50 579 523]{kaonpionT1u.eps}
\includegraphics[width=2.2in,bb=7 50 579 523]{localpipi1.eps}
\includegraphics[width=2.2in,bb=7 50 579 523]{localpipi2.eps}
\end{center}
\caption{
(Left) Effective masses, $m_{\rm eff}(t)$ using $dt=3$, associated with a 
two-meson operator in the $T_{1u}$ irrep, having total isospin $I=\frac{1}{2}$ and 
zero total momentum, constructed from single-site kaon and pion operators having 
equal and opposite on-axis momenta of minimal nonzero magnitude.  Results on the 
$(24^3\vert 390)$ and $(32^3\vert 240)$ ensembles are shown.  The energies
of a free $\pi$ plus a free $K$ are indicated by horizontal dashed lines.  
(Center) Effective mass for one of our $I=1$ $\pi(1)\pi(-1)$ operators in the 
$T_{1u}^+$ channel, consisting of single-site pion operators having equal and 
opposite on-axis momenta of minimal nonzero magnitude, compared to the effective 
mass of a localized $\pi\pi$ operator, described in Eq.~(\ref{eq:localpipiT1um}),
on the $(24^3\vert 390)$ ensemble.  (Right) Effective mass for one of our 
$I=2$ $\pi(0)\pi(0)$ operators in the $A_{1g}^+$ channel, consisting of single-site 
pion operators each having zero momenta, compared to the effective mass of a 
localized $\pi\pi$ operator, described in Eq.~(\ref{eq:localpipiA1gp}), on 
the $(24^3\vert 390)$ ensemble. 
\label{fig:kaonpion}}
\end{figure*}

In addition to efficiency, there are good physical reasons for using
such multi-hadron operators.  Hadron-hadron interactions in finite volume
move the energies of any two-hadron systems away from their 
free two-particle energies, and the interacting two-particle states
could involve distributions of different relative momenta.  
However, such interactions are usually small and the relative momenta
used in our operators should presumably dominate in most cases.  
Also, we will always utilize multi-hadron operators with a variety of
different relative momenta to accommodate the effects of such interactions.
The performance of some of our $\pi\pi$ operators are compared to
localized multi-hadron operators in Fig.~\ref{fig:kaonpion}, discussed below.

Each single-hadron operator is labelled by total isospin $I$, the projection
of the total isospin $I_3$, strangeness $S$, three-momentum $\pvec$,
the little group irrep $\Lambda$, the row of the irrep $\lambda$,
and $i$, which denotes all other identifying information, such as
the displacement type and index.  Hence, basis operators for the
two-hadron operators can be written
\begin{equation}
   B^{I_aI_{3a}S_a}_{\pvec_a\Lambda_a\lambda_a i_a}
  \  B^{I_bI_{3b}S_b}_{\pvec_b\Lambda_b\lambda_b i_b},
\end{equation}
where $B$ denotes either a baryon (as a Grassmann operator) or a
meson, and $a$ and $b$ denote the separate hadrons.  Although the above operators 
form a perfectly acceptable basis of operators, they do not transform irreducibly 
under isospin rotations, nor under the $O_h^1$ symmetry transformations.

To construct such irreducible operators, we first need to know how the
basis operators transform.  Eqs.~(\ref{eq:hadrontransform}), (\ref{eq:wignerrot}), 
(\ref{eq:isospindef}), and (\ref{eq:mesontransformG}) summarize the
important transformation properties of our single-hadron operators.
Starting with the basis operators above, we first identify subsets of these operators
that transform among themselves under all $O_h^1$ transformations.
The total momentum $\pvec=\pvec_a+\pvec_b$ is the first quantity we consider.
We identify the little group of transformations that leave $\pvec$ invariant,
then our goal is to construct operators that transform according to the $\Lambda$ 
irrep of that little group.  Under group element $R$ of $O_h^1$, the single particle 
operator $B^{I_aI_{3a}S_a}_{\pvec_a\Lambda_a\lambda_a i_a}$ transforms into
one that has possibly a different momentum $R\pvec_a$ and is a linear
combination of the different rows of the $\Lambda_a$ irrep. The following quantities 
do not change under the $O_h^1$ transformations:
$\Lambda_a,\ \Lambda_b,\ i_a,\ i_b$.
We fix the above quantities, then apply the same group-theoretical projections
as for the single-hadrons to construct the linear combinations that transform 
irreducibly.  This can be easily done since we know exactly how each of the
basis operators transforms under any $O_h^1$ transformation.  Next, we 
form flavor combinations that transform irreducibly under isospin rotations.
Lastly, we apply $G$-parity projections, whenever suitable.

The above procedure is applied for two-particle operators having total momenta in the
directions of the reference momenta $\bm{p}_{\rm ref}$ in Table~\ref{tab:refmoms}. 
An operator having total momentum in any other direction is obtained by applying
the appropriate reference rotation $R_{\rm ref}$ in Table~\ref{tab:refmoms} to the
appropriate operator having momentum in a reference direction.

\begin{figure*}
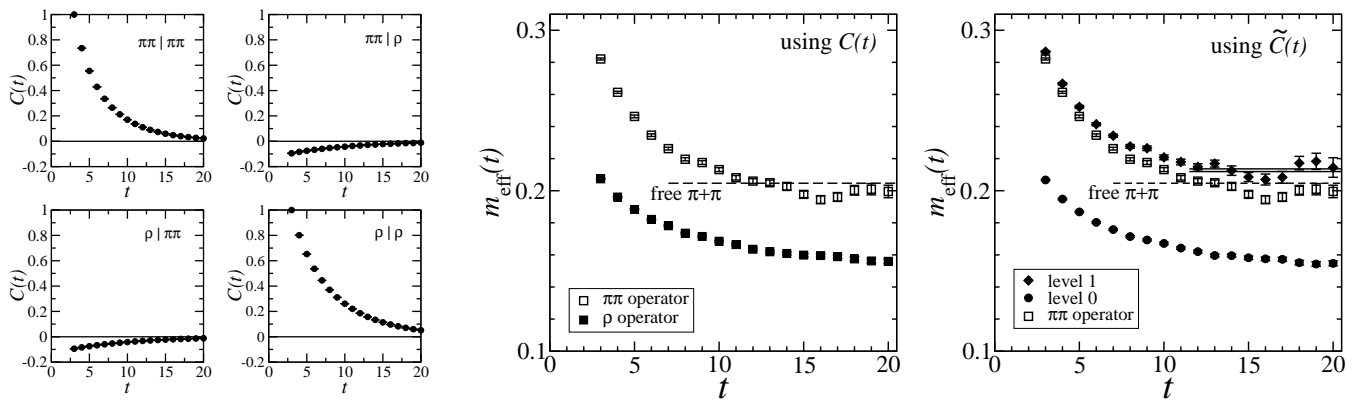

\begin{center}
\begin{minipage}{7.0in}
\includegraphics[width=2.2in,bb=2 8 582 549]{rho-pipi0.eps}\qquad
\includegraphics[width=2.2in,bb=7 30 541 523]{rho-pipi1.eps}\quad
\includegraphics[width=2.2in,bb=7 30 541 523]{rho-pipi2.eps}\quad
\end{minipage}
\end{center}
\caption{
Mixing of a quark-antiquark single-site $\rho$ operator and a two-pion operator.
Both operators transform according to the $T_{1u}^+$ irrep and create states of 
zero total momentum.  In the two-pion operator, each pion is a single-site operator 
travelling with minimal nonzero on-axis momenta.  (Left) The $2\times 2$ correlation 
matrix $C^\prime_{ij}(t)=C_{ij}(t)\ (\ C_{ii}(\tau_N)C_{jj}(\tau_N)\ )^{-1/2}$ with 
$\tau_N=3$ in terms of temporal lattice spacing $a_t$. (Center)
Effective masses, $m_{\rm eff}(t)$ using $dt=3$, associated with the diagonal
elements of the original correlator matrix $C(t)$.  (Right)  Effective masses 
associated with the diagonal elements of the rotated correlator 
$\widetilde{C}(t)$ defined in 
Eq.~\ref{eq:tildecor}, compared to the effective mass (hollow squares) of the 
$\pi\pi$ operator as also shown in the center plot.  The horizontal dashed
lines show the location of the free $\pi+\pi$ energy.  The horizontal solid lines
show the fit value of the energy of the interacting $\pi\pi$ state.
These results were obtained on the $(24^3\vert 390)$ ensemble. 
\label{fig:rhopipi}}
\end{figure*}

\begin{figure*}
\begin{center}
\begin{minipage}{7.0in}
\includegraphics[width=6.5in,bb=29 61 927 745]{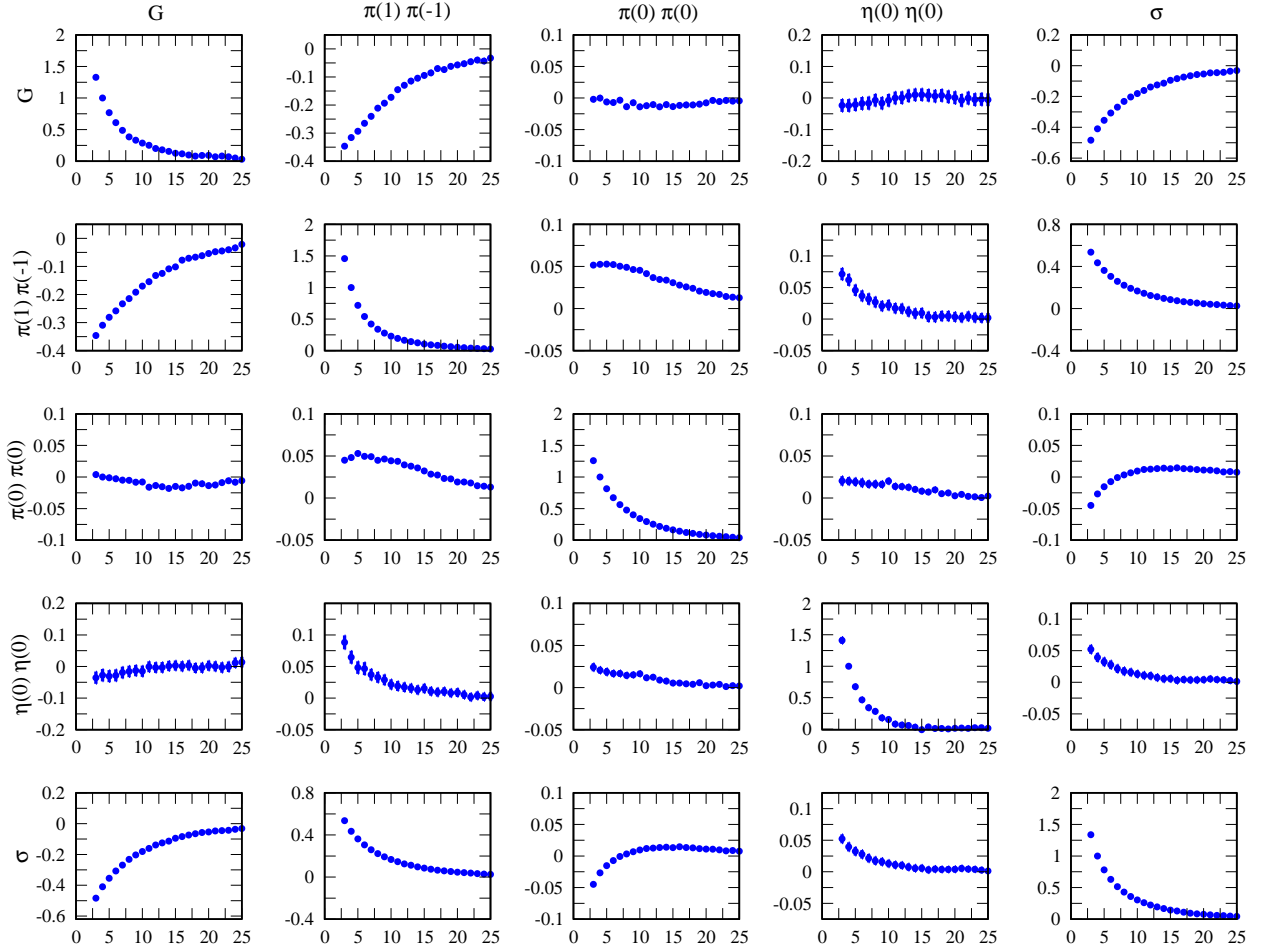}\qquad
\end{minipage}
\end{center}
\caption{
The $5\times 5$ matrix $C^\prime_{ij}(t)=C_{ij}(t)\ (\ C_{ii}(\tau_N)C_{jj}(\tau_N)\ )^{-1/2}$, 
with $\tau_N=4$, of the temporal correlations of five isoscalar operators in the 
scalar $A_{1g}^+$ sector for the $(24^3\vert 390)$ ensemble.  
$G$ is the $G_\Delta$ glueball operator described in Sec.~\ref{sec:glueball},
and $\sigma$ refers to a single-site quark-antiquark operator.  Three of the
operators are $I=0$ two-meson operators constructed out of single-site single-meson 
operators having equal and opposite momenta. In the $\eta(0)\eta(0)$ and 
$\pi(0)\pi(0)$ operators, each meson has zero momentum, whereas each pion
in the $\pi(1)\pi(-1)$ has minimal nonzero on-axis momentum.
Note that large vacuum expectation values have been subtracted to obtain
each of these correlators.
\label{fig:isoscalarA1g}}
\end{figure*}

In order to test the effectiveness of the two-hadron operators
that we have designed, we examined the effective masses associated with
the correlators of a variety of two-hadron operators.  We also evaluated
several correlation matrices mixing single and two-hadron operators.
Samples of our results are shown in 
Figs.~\ref{fig:kaonpion}, \ref{fig:rhopipi}, and \ref{fig:isoscalarA1g}.

In the left plot of Fig.~\ref{fig:kaonpion}, effective masses using $dt=3$ 
associated with a two-meson operator in the $T_{1u}$ irrep are shown.  
The two-meson operator has total isospin $I=\frac{1}{2}$ and zero total 
momentum and is constructed from single-site kaon and pion operators having 
equal and opposite on-axis momenta of minimal nonzero magnitude.  Results 
on the $(24^3\vert 390)$ and $(32^3\vert 240)$ ensembles are shown
and compared to the energies of a free $\pi$ plus a free $K$, indicated by
horizontal dashed lines.

An alternative design for a two-hadron operator is to use a suitable
localized operator.  For example, localized $\pi\pi$ operators in the
$I=2, A_{1g}^+$ and $I=1, T_{1u}^+$ channels can be obtained using
\begin{eqnarray}
 (\pi\pi)^{A_{1g}^+}(t) &=& \sum_{\bm{x}}
  \pi^+(\bm{x},t)\ \pi^+(\bm{x},t),
 \label{eq:localpipiA1gp}\\
 (\pi\pi)^{T_{1u}^+}(t) &=& \!\!\!\!\!\sum_{\bm{x},k=1,2,3}
 \!\!\!\!\Bigl\{ \pi^+(\bm{x},t)\ \Delta_k\pi^0(\bm{x},t) \nonumber\\
 && \qquad-\pi^0(\bm{x},t)\ \Delta_k\pi^+(\bm{x},t)\Bigr\},
 \label{eq:localpipiT1um}
\end{eqnarray}
where $\pi(\bm{x},t)$ is a single-site pion field using a standard
$\gamma_5$ construction with the LapH-smeared quark fields, and
$\Delta_k\pi(\bm{x},t)=\pi(\bm{x}\!+\!\widehat{\bm{k}},t)
 -\pi(\bm{x}\!-\!\widehat{\bm{k}},t)$.  The superscripts indicate
the electric charges associated with each field. In such localized $\pi\pi$
operators, the individual pions do not have definite momenta.

The center and right plots of Fig.~\ref{fig:kaonpion} compare the
effective masses for our $\pi\pi$ operators to those for these
localized $\pi\pi$ operators.  The center plot of Fig.~\ref{fig:kaonpion}
shows the $dt=3$ effective mass for one of our $I=1$ $\pi(1)\pi(-1)$ 
operators in the $T_{1u}^+$ channel, consisting of single-site pion operators 
having equal and opposite on-axis momenta of minimal nonzero magnitude, 
compared to the effective mass of the localized $\pi\pi$ operator, given 
in Eq.~(\ref{eq:localpipiT1um}), on the $(24^3\vert 390)$ ensemble.  The
right plot of Fig.~\ref{fig:kaonpion} shows the effective mass for one of our 
$I=2$ $\pi(0)\pi(0)$ operators in the $A_{1g}^+$ channel, consisting of 
single-site pion operators each having zero momenta, compared to the 
effective mass of the localized $\pi\pi$ operator, given in 
Eq.~(\ref{eq:localpipiA1gp}), on the $(24^3\vert 390)$ ensemble. 
One sees that the effective masses of the localized $\pi\pi$ operators
lie well above those of our operators, indicating that they contain much 
more excited-state contamination.  These effective masses are compared
to the energies of the ground state $\rho$ and the free $\pi+\pi$ energies,
indicated by horizontal dashed lines, in this figure.  Note that, in
addition to having much less excited-state contamination, the two-pion 
operators comprised of individual pions having definite momenta are
also much easier to make in large numbers, compared to the localized
multi-hadron operators.

The mixing of a single-site $\rho$ quark-antiquark operator and a two-pion operator
is examined in Fig.~\ref{fig:rhopipi}.  Both operators transform according to the 
$T_{1u}^+$ irrep and create states of zero total momentum.  In the two-pion operator, 
each pion is a single-site operator travelling with minimal nonzero on-axis momenta.  
In the left plot, the $2\times 2$ real and symmetric correlation matrix 
$C^\prime_{ij}(t)=C_{ij}(t)\ (\ C_{ii}(\tau_N)C_{jj}(\tau_N)\ )^{-1/2}$, with 
$\tau_N=3$, is shown.  Irrelevant normalization factors are removed by 
dividing the original correlation matrix by $(\ C_{ii}(\tau_N)C_{jj}(\tau_N)\ )^{1/2}$
for some early $\tau_N$.  The condition number of this matrix $C^\prime(t)$ at time
$t=4$ is 1.21, and at $t=20$, the condition number is 3.05.
In each matrix element label, the symbol to the right
of the vertical bar indicates the source operator and the symbol to the left
of the vertical bar denotes the sink operator.  These results were obtained on the 
$(24^3\vert 390)$ ensemble.  This figure provides further evidence that the
stochastic LapH method is well suited to studying systems involving the
mixing of single and multi-hadron operators, and that our
smearing parameters and noise dilution schemes were chosen judiciously.
One sees that there is a small but noticeably nonzero mixing between these
operators.

In the center plot of Fig.~\ref{fig:rhopipi}, the effective masses associated with 
the diagonal elements of the original correlator matrix $C(t)$ are displayed.  
A rotated correlation matrix $\widetilde{C}(t)$ can be defined by 
first solving for the unitary matrix $U$ in the eigenvalue equation below:
\begin{equation}
 C(\tau_0)^{-1/2}\ C(\tau_D)\ C(\tau_0)^{-1/2} = U\ \widetilde{C}(\tau_D)\ U^\dagger,
\end{equation}
where $\widetilde{C}(\tau_D)$ is a diagonal matrix and $C(t)$ is the original
Hermitian correlation matrix, then writing
\begin{equation}
  \widetilde{C}(t)=U^\dagger C(\tau_0)^{-1/2}\ C(t)\ C(\tau_0)^{-1/2}\ U.
\label{eq:tildecor}
\end{equation}
By construction, $\widetilde{C}(\tau_0)$ is the identity matrix, and 
$\widetilde{C}(\tau_D)$ is a diagonal matrix.  For all other times, $\widetilde{C}(t)$
need not be diagonal.  Note that $\widetilde{C}(t)$ is not a principal
correlator, such as used in Ref.~\cite{Blossier:2009kd}, since the diagonalization 
is not done for every time.  We use $\tau_0=8$ and $\tau_D=15$, and we find that 
$\widetilde{C}(t)$ remains diagonal, within statistical precision, for all $t$ 
exceeding $\tau_D$.  In the right plot in Fig.~\ref{fig:rhopipi}, the effective 
masses associated with the diagonal elements of $\widetilde{C}(t)$ are shown, 
and are labeled by level 0 and level 1.

The purpose of Fig.~\ref{fig:rhopipi} is to demonstrate the effectiveness
of the stochastic LapH method in providing estimates of correlation matrices 
involving both single and two-hadron operators that are accurate enough to allow 
reliable diagonalizations and extractions of excited-state energies.  Here, it 
is not our intent to carry out a detailed analysis of this correlation matrix.  
In future work, we shall include many more operators and attempt to extract the 
energies of a larger number of low-lying stationary states.  For now, we only
make a few remarks concerning this correlation matrix.  

In the center plot, one sees that the effective mass associated with the original 
$\rho$ operator tends to the lowest-lying stationary-state energy in this channel, 
consistent with expectations.  The effective mass associated with the two-pion 
operator nearly levels off at a higher-lying energy, but it will eventually
fall to the lowest-lying $\rho$ energy, given large enough $t$.  The coupling 
of our two-pion operator to the $\rho$ state is nonzero, but apparently 
much smaller than its coupling to the lowest-lying two-pion stationary state.
The first excited-state energy can be revealed by constructing the
rotated correlator $\widetilde{C}(t)$ described above.  By ensuring the
off-diagonal elements of $\widetilde{C}(t)$ are zero within statistical
precision for all $t>\tau_D$, the effective masses associated with the diagonal 
elements of $\widetilde{C}(t)$ tend to the two lowest-lying stationary-state 
energies in this symmetry channel, as shown in the right plot of 
Fig.~\ref{fig:rhopipi}.  The solid horizontal lines indicate the location of the 
first excited-state energy (level 1) as determined by a single-exponential fit, 
with wrap-around, to $\widetilde{C}_{11}(t)$. An excellent fit quality is obtained,
and the uncertainty in the energy from the fit is indicated by the two parallel 
lines.  This energy is compared to the free two-pion energy, indicated by the
dashed horizontal line.  The effective mass for the original two-pion operator 
is also shown in the right plot (hollow squares) for comparison.  The 
diagonalization appears to remove the small coupling to the $\rho$ stationary 
state.  

The effective mass associated with $\widetilde{C}_{00}(t)$ for level 0 
shows a very slight downward drift at large $t$.  This is due to a temporal 
wrap-around effect in which our operators create a $\pi\pi$ state and one 
$\pi$ propagates forward in time while the other $\pi$ propagates backwards 
in time, producing a small contribution that is essentially constant with
respect to time.  A fit to this correlator using $A(e^{-Et}+e^{-E(N_t-t)})$,
which ignores this contribution, for times $t=17-25$ produces a poor 
$\chi^2$/dof = 2.51, whereas a fit including a constant term 
$A(e^{-Et}+e^{-E(N_t-t)})+B$ for $t=17-25$ produces a good
fit quality $\chi^2$/dof = 1.03, with a fit value for the energy $E=0.1638(20)$.

A last example of the effectiveness of the stochastic LapH method in dealing 
with mixings between single and two-meson operators is given in 
Fig.~\ref{fig:isoscalarA1g}, which shows a rescaled $5\times 5$ correlation 
matrix $C^\prime_{ij}(t)=C_{ij}(t)\ (\ C_{ii}(\tau_N)C_{jj}(\tau_N)\ )^{-1/2}$, 
with $\tau_N=4$, of the temporal correlations of five isoscalar operators in 
the scalar $A_{1g}^+$ sector for the $(24^3\vert 390)$ ensemble.  
The condition number of $C^\prime(3)$ is 4.82, of $C^\prime(8)$ is
18.1, and of $C^\prime(12)$ is 57.7.
$G$ is the $G_\Delta$ glueball operator described in Sec.~\ref{sec:glueball},
and $\sigma$ refers to a single-site quark-antiquark operator.  Three of the
operators are $I=0$ two-meson operators constructed out of single-site single-meson 
operators having equal and opposite momenta. In the $\eta(0)\eta(0)$ and 
$\pi(0)\pi(0)$ operators, each meson has zero momentum, whereas each pion
in the $\pi(1)\pi(-1)$ has minimal nonzero on-axis momentum.
In this channel, the correlation matrix elements are defined, 
for $N_t\rightarrow\infty$, by
%\begin{eqnarray}
%    C_{ij}(t_F-t_0) &= &
%  \langle 0\vert\ O_i(t_F)\ \overline{O}_j(t_0)\ \vert 0\rangle\nonumber\\
%  &-& \langle 0\vert\ O_i(t_F)\ \vert 0\rangle\langle 0\vert
%\ \overline{O}_j(t_0)\ \vert 0\rangle.
%\end{eqnarray}
\begin{equation}
    C_{ij}(t) =
  \langle O_i(t)\ \overline{O}_j(0)\rangle
   - \langle O_i\rangle\langle \overline{O}_j\rangle.
\end{equation}
Large vacuum expectation values have been subtracted to obtain
each of these correlators.  Let $O_0, O_1, O_2, O_3, O_4$ denote the 
glueball operator $G_\Delta$, the $\pi(1)\pi(-1)$, the $\pi(0)\pi(0)$, 
the $\eta(0)\eta(0)$, and the $\sigma$ operators, respectively.  The 
ratios of the diagonal elements of the correlation matrix at time 
separation $t=3$ over the squares of their respective 
vacuum expectation values for these five operators are shown below:
\begin{eqnarray*}
 C_{00}(3)/\langle O_0\rangle^2 &=& 0.00001205(29),\\
 C_{11}(3)/\langle O_1\rangle^2 &=& 0.006657(32),\\
 C_{22}(3)/\langle O_2\rangle^2 &=& 0.11396(74),\\
 C_{33}(3)/\langle O_3\rangle^2 &=& 0.155(12),\\
 C_{44}(3)/\langle O_4\rangle^2 &=& 0.002371(12).
\end{eqnarray*}
The smallness of these numbers demonstrates the largeness of the
vacuum expectation values for these correlators.  This is a notoriously 
difficult channel to study, but the stochastic LapH method appears to 
produce results of adequate precision, even for the 
$\eta\eta\rightarrow\eta\eta$ correlator, which includes many diagrams 
involving internal quark loops and other same-time quark lines.  Additional 
operators, such as $\overline{K}K$ and other $\pi\pi, \eta\eta$, operators 
are needed to reliably study the physics here.  We certainly plan to 
investigate this channel in much more detail in the future using a larger 
number of operators.

Table~\ref{tab:flavsectors} summarizes the different flavor types of single
and two-hadron operators that we plan to include in our first survey of
the spectrum of stationary-state energies.  We plan to study all
bosonic and fermionic flavor sectors involving the $u,d,s$ quarks that involve
up to two meson and meson-baryon pairs.  The particle types to be studied in each 
flavor sector are 
listed in this table.  $G$ denotes a glueball operator.  In this table, $\pi$ 
refers to \textit{any} isovector quark-antiquark meson operator with flavor content 
such as $\overline{d}u$ (such as $a$, $b$, $\pi$, and $\rho$ mesons), $\eta$ refers 
to any $\overline{u}u+\overline{d}d$ isoscalar meson operator, $\phi$ refers to any 
$\overline{s}s$ isoscalar meson operator,  $K$ refers to any quark-antiquark operator 
having flavor content $\overline{s}u$ or $\overline{s}d$ such that its strangeness 
is $S=1$, and $\overline{K}$ is any quark-antiquark operator having flavor content 
$\overline{u}s$ or $\overline{d}s$ such that its strangeness is $S=-1$. 
Our current plans do not include states containing two or more baryons or 
three or more mesons, although the computational technology can easily accommodate 
such states.

\begin{table}[t]
\caption[flavbos]{
The following bosonic and fermionic flavor sectors involving only 
the $u,d,s$ quarks will be studied.  $J$ denotes total spin, $I$ denotes total 
isospin, and $S$ is the total strangeness.  The particle contents to be studied 
in each flavor sector are listed.  $G$ denotes a glueball operator.  Here, $\pi$ 
refers to \textit{any} isovector quark-antiquark meson operator with flavor content 
such as $\overline{d}u$ (such as $a$, $b$, $\pi$, and $\rho$ mesons), $\eta$ refers 
to any $\overline{u}u+\overline{d}d$ isoscalar meson operator, $\phi$ refers to any 
$\overline{s}s$ isoscalar meson operator,  $K$ refers to any quark-antiquark operator 
having flavor content $\overline{s}u$ or $\overline{s}d$ such that its strangeness 
is $S=1$, and $\overline{K}$ is any quark-antiquark operator having flavor content 
$\overline{u}s$ or $\overline{d}s$ such that its strangeness is $S=-1$. 
\label{tab:flavsectors}}
\begin{ruledtabular}
\begin{tabular}{ccrl}
$(-1)^{2J}$ & $I$ & $S$ & Particle content \\ \hline
 1 & 0 & 0 & $\eta,\ \phi,\ G,\ \eta\eta,\ \eta\phi,\ \phi\phi,\ \pi\pi,\ \overline{K}K$  \\
 1 & 1 & 0 & $\pi,\ \pi\pi,\ \eta\pi,\ \phi\pi,\ \overline{K}K$          \\
 1 & 2 & 0 & $\pi\pi$        \\
 1 & $\frac{1}{2}$ & $1$ & $K,\ K\pi,\ K\eta,\ K\phi$  \\
 1 & $\frac{3}{2}$ & $1$ & $K\pi$  \\
 1 & 0 & $2$ &   $KK$ \\
 1 & 1 & $2$ &   $KK$  \\ \hline
 $-1$ & $\frac{1}{2}$ & 0 & $N,\ N\eta,\ N\phi,\ N\pi,\ \Delta\pi,\ \Lambda K,\ \Sigma K$  \\
 $-1$ &  $\frac{3}{2}$ & 0 & $\Delta,\ \Delta\eta,\ \Delta\phi,\ \Delta\pi,\ N\pi,\ \Sigma K$  \\
 $-1$ &  $\frac{5}{2}$ & 0 & $\Delta\pi$        \\
 $-1$ &  0 & $-1$ & $\Lambda,\ \Lambda\eta,\ \Lambda\phi,\ N\overline{K},\ \Sigma\pi,\ \Xi K$  \\
 $-1$ &  1 & $-1$ & $\Sigma,\ \Sigma\eta,\ \Sigma\phi,\ \Sigma\pi,\ N\overline{K},
  \ \Delta\overline{K},\ \Lambda\pi,\ \Xi K$  \\
 $-1$ &  2 & $-1$ & $\Delta\overline{K},\ \Sigma\pi$  \\
 $-1$ &  $\frac{1}{2}$ & $-2$ & $\Xi,\ \Lambda\overline{K},\ \Xi\eta,\ \Xi\phi,\ \Xi\pi,
  \ \Sigma\overline{K},\ \Omega K$  \\
 $-1$ &   $\frac{3}{2}$ & $-2$ & $\Sigma\overline{K},\ \Xi\pi$          \\
 $-1$ &  0 & $-3$ &   $\Omega,\ \Xi\overline{K},\ \Omega\eta,\ \Omega\phi$ \\
 $-1$ &  1 & $-3$ &   $\Omega\pi,\ \Xi\overline{K}$  
\end{tabular}
\end{ruledtabular}
\end{table}

\section{A new glueball operator}
\label{sec:glueball}

\begin{figure*}
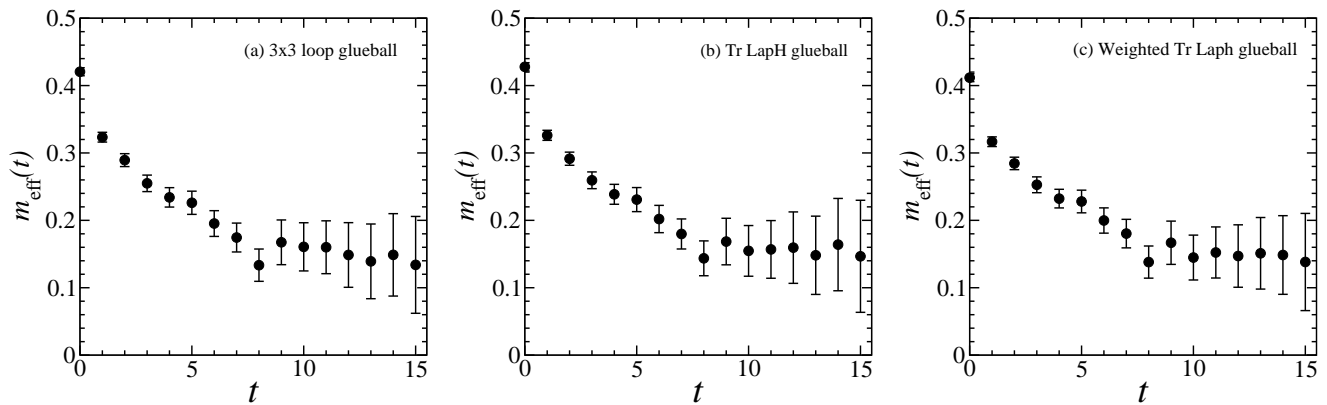

\begin{center}
\begin{minipage}{7.0in}
\includegraphics[width=2.2in,bb=7 50 541 533]{glueball1.eps}\quad
\includegraphics[width=2.2in,bb=7 50 541 533]{glueball2.eps}\quad
\includegraphics[width=2.2in,bb=7 50 541 533]{glueball3.eps}\\[-8mm]
\end{minipage}
\end{center}
\caption{
Comparison of the effective masses, $m_{\rm eff}(t)$ using $dt=3$, associated with 
three different scalar glueball operators on the $(24^3\vert 390)$ ensemble.
(a) The leftmost plot shows the effective mass using an operator defined by a sum 
of $3\times 3$ loops of the smeared gauge link variables that is rotationally and 
translationally invariant. (b) The middle plot uses the new TrLapH glueball 
operator $G_\Delta$, defined in Eq.~(\ref{eq:Gdelta}). (c) The rightmost plot shows 
the results using the weighted TrLapH glueball operator $G_W$, defined in 
Eq.~(\ref{eq:GWdelta}).  One observes very little difference between these plots, 
suggesting that these operators are comparable in usefulness for studying the 
scalar glueball.  Each effective mass eventually tends toward the energy of two 
pions at rest, demonstrating non-negligible coupling of these operators to
$\pi\pi$ states.
\label{fig:glueballops}}
\end{figure*}

Determining stationary-state energies in the interesting scalar isoscalar sector
will ultimately involve including a scalar glueball operator.  Glueballs
are hypothetical particles comprised predominantly of gluons, having no valence
quarks.  Scalar glueball operators are usually constructed using a sum of 
gauge-invariant loops of the smeared spatial link variables on a single time slice, 
which is invariant under translations, rotations, and charge conjugation.  
However, any purely gluonic quantity with similar symmetry properties could 
presumably be used.  LapH quark-field smearing involves the covariant spatial 
Laplacian $\widetilde{\Delta}$.  The eigenvalues of the Laplacian
are invariant under rotations and gauge transformations, and so, are appropriate
for a scalar glueball operator.  The lowest-lying eigenvalue was studied,
as well as other functions of the eigenvalues.  We found that essentially any
combination of the low-lying eigenvalues worked equally well for
studying the scalar glueball.  Two operators in particular that we studied are
\begin{eqnarray}
 G_\Delta(t) &=& 
-{\rm Tr}[\Theta(\sigma_s^2+\widetilde{\Delta})\ \widetilde{\Delta}],
\label{eq:Gdelta}\\
 G_W(t) &=& -{\rm Tr}[\Theta(\sigma_s^2+\widetilde{\Delta})
 \ \widetilde{\Delta}\,\exp(-W\widetilde{\Delta}^2)].
\label{eq:GWdelta}
\end{eqnarray}
The first operator $G_\Delta$ in Eq.~(\ref{eq:Gdelta}), which we call the TrLapH 
operator, is perhaps the simplest operator that one can construct using the 
eigenvalues of the covariant Laplacian.  In the so-called weighted TrLapH operator
$G_W$ in Eq.~(\ref{eq:GWdelta}), we used $W=64$ in order that only a handful of the 
lowest-lying eigenvalues contribute.  

The effective masses obtained using $G_\Delta$ and $G_W$ on the $24^3\times 128$
ensemble are shown in Fig.~\ref{fig:glueballops}.  These effective masses
are compared to that obtained using a standard glueball operator which is a
sum of $3\times 3$ loops of the smeared gauge link variables that is rotationally 
and translationally invariant.  One observes very little difference between these 
effective masses, suggesting these operators are comparable in usefulness for 
studying the scalar glueball.  Similar conclusions were reached using
the $32^3\times 256$ lattice.  Each effective mass eventually tends toward 
the energy of two pions at rest, demonstrating non-negligible coupling of these 
operators to $\pi\pi$ states.  Thus, we plan to use the simplest operator
$G_\Delta(t)$ in future studies involving the scalar glueball.

\section{Conclusion}
\label{sec:conclude}

Multi-hadron operators are crucial for reliably extracting the masses of 
excited states lying above multi-hadron thresholds in lattice QCD Monte Carlo 
calculations. Multi-hadron operators with significant coupling to the low-lying 
multi-hadron states of interest can be obtained by combining single-hadron 
operators of various momenta.  The construction and testing of single-hadron
operators of definite momentum, and their combinations into two-hadron
operators was the main subject of this work.

The approach of Ref.~\cite{baryons2005A} was extended to meson operators of 
zero momentum, and to both meson and baryon operators having definite 
nonzero momentum.  Our operator design utilizes group-theoretical projections.
The point and space groups we use are well known, and the properties of their 
irreducible representations are widely available in the literature. However,
we collected together and presented in this paper some of the specific group 
theory details needed for our operator construction for the convenience of the
reader and to explicitly state our conventions and the notations we use.

Tests of our single-hadron operators using a stochastic 
method of treating the low-lying modes of quark propagation which exploits 
Laplacian Heaviside quark-field smearing were presented.  These tests were
carried out on $24^3\times 128$ and $32^3\times 256$ anisotropic lattices with 
pion masses $m_\pi\approx 390$ and 240~MeV.  A new glueball operator was also 
introduced and tested.  We demonstrated that computing the mixing of this 
glueball operator with a quark-antiquark operator, $\pi\pi$, and $\eta\eta$ 
operators is feasible with the stochastic LapH method.

The stochastic LapH method provides reliable estimates of all temporal 
correlations that will be needed for a comprehensive survey of the low-lying 
spectrum of QCD stationary states in finite volume.  The method works well
even for those correlators that are particularly difficult to 
compute, such as $\eta\eta\rightarrow\eta\eta$ in the scalar channel, which
involves the subtraction of a large vacuum expectation value. 
The effectiveness of the method can be traced to two of its key features: the
use of noise dilution projectors that interlace in time, and the use of
$Z_N$ noise in the subspace defined by the Laplacian Heaviside quark-field
smearing.  Introducing noise in the LapH subspace results in greatly
reduced variances in temporal correlations compared to methods that 
introduce noise on the entire lattice.  Although the number of Laplacian
eigenvectors needed to span the LapH subspace rises dramatically with the
spatial volume, the number of inversions of the Dirac matrix
needed for a target accuracy is remarkably insensitive to the lattice volume,
once a sufficient number of dilution projectors is 
introduced\cite{Morningstar:2011ka}.

In addition to increased efficiency, the stochastic LapH method has other
advantages.  The method leads to complete factorization of hadron 
sources and sinks in temporal correlations, which greatly simplifies 
the logistics of evaluating correlation matrices involving large numbers
of operators.  Implementing the Wick contractions of the quark lines is
also straightforward.  Contributions from different Wick orderings 
within a class of quark-line diagrams differ only by permutations of the
noises at the source. 

In the future, we plan to carry out a comprehensive survey of the
excitation spectrum of the stationary states of QCD involving mesons
and baryons containing $u,d,s$ quarks.  Various scattering phase shifts
and decay constants will also be investigated.  The needed
single-meson and single-baryon sources and sinks for a large number
of different momenta have been computed and stored for three Monte Carlo
ensembles.  The development and testing of the software to combine these 
sources and sinks via Wick contractions of the quark fields into 
temporal correlators has been completed, and the final stages of our
operator selections are now in progress.  Results for the
spectrum, involving both single- and two-hadron operators and using the
technology described in this work, will appear in future publications.

\begin{acknowledgments}
This work was supported by the U.S.~National Science Foundation 
under awards PHY-0510020, PHY-0653315, PHY-0704171, PHY-0969863, and
PHY-0970137, and through TeraGrid/XSEDE resources provided by the 
Pittsburgh Supercomputer Center, the Texas Advanced Computing Center, 
and the National Institute for Computational
Sciences under grant numbers TG-PHY100027 and TG-MCA075017.  
The USQCD QDP++ library\cite{Edwards:2004sx} and the Improved
BiCGStab solver in Chroma were used in developing 
the software for the calculations reported here.   We acknowledge
conversations with Balint Joo, Christian Lang, Daniel Mohler, 
Mike Peardon, Sasa Prelovsek, and Christopher Thomas.
\end{acknowledgments}

\bibliography{cited_refs}

\begin{thebibliography}{33}
\expandafter\ifx\csname natexlab\endcsname\relax\def\natexlab#1{#1}\fi
\expandafter\ifx\csname bibnamefont\endcsname\relax
  \def\bibnamefont#1{#1}\fi
\expandafter\ifx\csname bibfnamefont\endcsname\relax
  \def\bibfnamefont#1{#1}\fi
\expandafter\ifx\csname citenamefont\endcsname\relax
  \def\citenamefont#1{#1}\fi
\expandafter\ifx\csname url\endcsname\relax
  \def\url#1{\texttt{#1}}\fi
\expandafter\ifx\csname urlprefix\endcsname\relax\def\urlprefix{URL }\fi
\providecommand{\bibinfo}[2]{#2}
\providecommand{\eprint}[2][]{\url{#2}}

\bibitem[{\citenamefont{Basak et~al.}(2005{\natexlab{a}})\citenamefont{Basak,
  Edwards, Fleming, Heller, Morningstar, Richards, Sato, and
  Wallace}}]{baryons2005A}
\bibinfo{author}{\bibfnamefont{S.}~\bibnamefont{Basak}},
  \bibinfo{author}{\bibfnamefont{R.}~\bibnamefont{Edwards}},
  \bibinfo{author}{\bibfnamefont{G.}~\bibnamefont{Fleming}},
  \bibinfo{author}{\bibfnamefont{U.}~\bibnamefont{Heller}},
  \bibinfo{author}{\bibfnamefont{C.}~\bibnamefont{Morningstar}},
  \bibinfo{author}{\bibfnamefont{D.}~\bibnamefont{Richards}},
  \bibinfo{author}{\bibfnamefont{I.}~\bibnamefont{Sato}}, \bibnamefont{and}
  \bibinfo{author}{\bibfnamefont{S.}~\bibnamefont{Wallace}},
  \bibinfo{journal}{Phys. Rev. D} \textbf{\bibinfo{volume}{72}},
  \bibinfo{pages}{094506} (\bibinfo{year}{2005}{\natexlab{a}}).

\bibitem[{\citenamefont{Basak et~al.}(2005{\natexlab{b}})\citenamefont{Basak,
  Edwards, Fleming, Heller, Morningstar, Richards, Sato, and
  Wallace}}]{baryons2005B}
\bibinfo{author}{\bibfnamefont{S.}~\bibnamefont{Basak}},
  \bibinfo{author}{\bibfnamefont{R.}~\bibnamefont{Edwards}},
  \bibinfo{author}{\bibfnamefont{G.}~\bibnamefont{Fleming}},
  \bibinfo{author}{\bibfnamefont{U.}~\bibnamefont{Heller}},
  \bibinfo{author}{\bibfnamefont{C.}~\bibnamefont{Morningstar}},
  \bibinfo{author}{\bibfnamefont{D.}~\bibnamefont{Richards}},
  \bibinfo{author}{\bibfnamefont{I.}~\bibnamefont{Sato}}, \bibnamefont{and}
  \bibinfo{author}{\bibfnamefont{S.}~\bibnamefont{Wallace}},
  \bibinfo{journal}{Phys. Rev. D} \textbf{\bibinfo{volume}{72}},
  \bibinfo{pages}{074501} (\bibinfo{year}{2005}{\natexlab{b}}).

\bibitem[{\citenamefont{Basak et~al.}(2007)\citenamefont{Basak, Edwards,
  Fleming, Juge, Lichtl, Morningstar, Richards, Sato, and
  Wallace}}]{baryon2007}
\bibinfo{author}{\bibfnamefont{S.}~\bibnamefont{Basak}},
  \bibinfo{author}{\bibfnamefont{R.}~\bibnamefont{Edwards}},
  \bibinfo{author}{\bibfnamefont{G.}~\bibnamefont{Fleming}},
  \bibinfo{author}{\bibfnamefont{K.}~\bibnamefont{Juge}},
  \bibinfo{author}{\bibfnamefont{A.}~\bibnamefont{Lichtl}},
  \bibinfo{author}{\bibfnamefont{C.}~\bibnamefont{Morningstar}},
  \bibinfo{author}{\bibfnamefont{D.}~\bibnamefont{Richards}},
  \bibinfo{author}{\bibfnamefont{I.}~\bibnamefont{Sato}}, \bibnamefont{and}
  \bibinfo{author}{\bibfnamefont{S.}~\bibnamefont{Wallace}},
  \bibinfo{journal}{Phys. Rev. D} \textbf{\bibinfo{volume}{76}},
  \bibinfo{pages}{074504} (\bibinfo{year}{2007}).

\bibitem[{\citenamefont{Bulava et~al.}(2009)}]{nucleon2009}
\bibinfo{author}{\bibfnamefont{J.}~\bibnamefont{Bulava}} \bibnamefont{et~al.},
  \bibinfo{journal}{Phys. Rev. D} \textbf{\bibinfo{volume}{79}},
  \bibinfo{pages}{034505} (\bibinfo{year}{2009}).

\bibitem[{\citenamefont{Morningstar et~al.}(2010)}]{Morningstar:2010ae}
\bibinfo{author}{\bibfnamefont{C.}~\bibnamefont{Morningstar}}
  \bibnamefont{et~al.}, \bibinfo{journal}{AIP Conf. Proc.}
  \textbf{\bibinfo{volume}{1257}}, \bibinfo{pages}{779} (\bibinfo{year}{2010}),
  \eprint{arXiv:1002.0818 [hep-lat]}.

\bibitem[{\citenamefont{Morningstar et~al.}(2011)\citenamefont{Morningstar,
  Bulava, Foley, Juge, Lenkner, Peardon, and Wong}}]{Morningstar:2011ka}
\bibinfo{author}{\bibfnamefont{C.}~\bibnamefont{Morningstar}},
  \bibinfo{author}{\bibfnamefont{J.}~\bibnamefont{Bulava}},
  \bibinfo{author}{\bibfnamefont{J.}~\bibnamefont{Foley}},
  \bibinfo{author}{\bibfnamefont{K.}~\bibnamefont{Juge}},
  \bibinfo{author}{\bibfnamefont{D.}~\bibnamefont{Lenkner}},
  \bibinfo{author}{\bibfnamefont{M.}~\bibnamefont{Peardon}}, \bibnamefont{and}
  \bibinfo{author}{\bibfnamefont{C.}~\bibnamefont{Wong}},
  \bibinfo{journal}{Phys. Rev. D} \textbf{\bibinfo{volume}{83}},
  \bibinfo{pages}{114505} (\bibinfo{year}{2011}).

\bibitem[{\citenamefont{Foley et~al.}(2010)\citenamefont{Foley, Wong
  et~al.}}]{Foley:2010vv}
\bibinfo{author}{\bibfnamefont{J.}~\bibnamefont{Foley}},
  \bibinfo{author}{\bibfnamefont{C.~H.} \bibnamefont{Wong}},
  \bibnamefont{et~al.}, \bibinfo{journal}{PoS}
  \textbf{\bibinfo{volume}{(LAT2010)}}, \bibinfo{pages}{098}
  (\bibinfo{year}{2010}), \eprint{arXiv:1011.0481 [hep-lat]}.

\bibitem[{\citenamefont{Bulava et~al.}(2010{\natexlab{a}})\citenamefont{Bulava,
  Foley et~al.}}]{Bulava:2010em}
\bibinfo{author}{\bibfnamefont{J.}~\bibnamefont{Bulava}},
  \bibinfo{author}{\bibfnamefont{J.}~\bibnamefont{Foley}},
  \bibnamefont{et~al.}, \bibinfo{journal}{PoS}
  \textbf{\bibinfo{volume}{(LAT2010)}}, \bibinfo{pages}{110}
  (\bibinfo{year}{2010}{\natexlab{a}}), \eprint{arXiv:1011.5277 [hep-lat]}.

\bibitem[{\citenamefont{Aoki et~al.}(2010)}]{Aoki:2009ix}
\bibinfo{author}{\bibfnamefont{S.}~\bibnamefont{Aoki}} \bibnamefont{et~al.},
  \bibinfo{journal}{Phys. Rev. D} \textbf{\bibinfo{volume}{81}},
  \bibinfo{pages}{074503} (\bibinfo{year}{2010}).

\bibitem[{\citenamefont{Mahbub et~al.}(2010)\citenamefont{Mahbub, Cais, Kamleh,
  Leinweber, and Williams}}]{Mahbub:2010jz}
\bibinfo{author}{\bibfnamefont{M.~S.} \bibnamefont{Mahbub}},
  \bibinfo{author}{\bibfnamefont{A.~O.} \bibnamefont{Cais}},
  \bibinfo{author}{\bibfnamefont{W.}~\bibnamefont{Kamleh}},
  \bibinfo{author}{\bibfnamefont{D.~B.} \bibnamefont{Leinweber}},
  \bibnamefont{and} \bibinfo{author}{\bibfnamefont{A.~G.}
  \bibnamefont{Williams}}, \bibinfo{journal}{Phys. Rev. D}
  \textbf{\bibinfo{volume}{82}}, \bibinfo{pages}{094504}
  (\bibinfo{year}{2010}).

\bibitem[{\citenamefont{Engel et~al.}(2010)\citenamefont{Engel, Lang, Limmer,
  Mohler, and Schafer}}]{Engel:2010my}
\bibinfo{author}{\bibfnamefont{G.~P.} \bibnamefont{Engel}},
  \bibinfo{author}{\bibfnamefont{C.~B.} \bibnamefont{Lang}},
  \bibinfo{author}{\bibfnamefont{M.}~\bibnamefont{Limmer}},
  \bibinfo{author}{\bibfnamefont{D.}~\bibnamefont{Mohler}}, \bibnamefont{and}
  \bibinfo{author}{\bibfnamefont{A.}~\bibnamefont{Schafer}},
  \bibinfo{journal}{Phys. Rev. D} \textbf{\bibinfo{volume}{82}},
  \bibinfo{pages}{034505} (\bibinfo{year}{2010}).

\bibitem[{\citenamefont{Bulava et~al.}(2010{\natexlab{b}})}]{Bulava:2010yg}
\bibinfo{author}{\bibfnamefont{J.}~\bibnamefont{Bulava}} \bibnamefont{et~al.},
  \bibinfo{journal}{Phys. Rev. D} \textbf{\bibinfo{volume}{82}},
  \bibinfo{pages}{014507} (\bibinfo{year}{2010}{\natexlab{b}}).

\bibitem[{\citenamefont{Dudek et~al.}(2011)}]{Peardon:2011tt}
\bibinfo{author}{\bibfnamefont{J.~J.} \bibnamefont{Dudek}}
  \bibnamefont{et~al.}, \bibinfo{journal}{Phys. Rev. D}
  \textbf{\bibinfo{volume}{83}}, \bibinfo{pages}{111502}
  (\bibinfo{year}{2011}).

\bibitem[{\citenamefont{Engel et~al.}(2012)\citenamefont{Engel, Lang, Limmer,
  Mohler, and Schafer}}]{Engel:2011aa}
\bibinfo{author}{\bibfnamefont{G.~P.} \bibnamefont{Engel}},
  \bibinfo{author}{\bibfnamefont{C.}~\bibnamefont{Lang}},
  \bibinfo{author}{\bibfnamefont{M.}~\bibnamefont{Limmer}},
  \bibinfo{author}{\bibfnamefont{D.}~\bibnamefont{Mohler}}, \bibnamefont{and}
  \bibinfo{author}{\bibfnamefont{A.}~\bibnamefont{Schafer}},
  \bibinfo{journal}{Phys. Rev. D} \textbf{\bibinfo{volume}{85}},
  \bibinfo{pages}{034508} (\bibinfo{year}{2012}).

\bibitem[{\citenamefont{Edwards et~al.}(2011)\citenamefont{Edwards, Dudek,
  Richards, and Wallace}}]{Edwards:2011jj}
\bibinfo{author}{\bibfnamefont{R.~G.} \bibnamefont{Edwards}},
  \bibinfo{author}{\bibfnamefont{J.~J.} \bibnamefont{Dudek}},
  \bibinfo{author}{\bibfnamefont{D.~G.} \bibnamefont{Richards}},
  \bibnamefont{and} \bibinfo{author}{\bibfnamefont{S.~J.}
  \bibnamefont{Wallace}}, \bibinfo{journal}{Phys. Rev. D}
  \textbf{\bibinfo{volume}{84}}, \bibinfo{pages}{074508}
  (\bibinfo{year}{2011}).

\bibitem[{\citenamefont{Mahbub et~al.}(2012)\citenamefont{Mahbub, Kamleh,
  Leinweber, Moran, and Williams}}]{Mahbub:2010rm}
\bibinfo{author}{\bibfnamefont{M.}~\bibnamefont{Mahbub}},
  \bibinfo{author}{\bibfnamefont{W.}~\bibnamefont{Kamleh}},
  \bibinfo{author}{\bibfnamefont{D.~B.} \bibnamefont{Leinweber}},
  \bibinfo{author}{\bibfnamefont{P.~J.} \bibnamefont{Moran}}, \bibnamefont{and}
  \bibinfo{author}{\bibfnamefont{A.~G.} \bibnamefont{Williams}},
  \bibinfo{journal}{Phys. Lett. B} \textbf{\bibinfo{volume}{707}},
  \bibinfo{pages}{389} (\bibinfo{year}{2012}).

\bibitem[{\citenamefont{Aoki et~al.}(2011)}]{Aoki:2011yj}
\bibinfo{author}{\bibfnamefont{S.}~\bibnamefont{Aoki}} \bibnamefont{et~al.},
  \bibinfo{journal}{Phys. Rev. D} \textbf{\bibinfo{volume}{84}},
  \bibinfo{pages}{094505} (\bibinfo{year}{2011}).

\bibitem[{\citenamefont{Lang et~al.}(2011)\citenamefont{Lang, Mohler,
  Prelovsek, and Vidmar}}]{Lang:2011mn}
\bibinfo{author}{\bibfnamefont{C.}~\bibnamefont{Lang}},
  \bibinfo{author}{\bibfnamefont{D.}~\bibnamefont{Mohler}},
  \bibinfo{author}{\bibfnamefont{S.}~\bibnamefont{Prelovsek}},
  \bibnamefont{and} \bibinfo{author}{\bibfnamefont{M.}~\bibnamefont{Vidmar}},
  \bibinfo{journal}{Phys. Rev. D} \textbf{\bibinfo{volume}{84}},
  \bibinfo{pages}{054503} (\bibinfo{year}{2011}).

\bibitem[{\citenamefont{Mohler et~al.}(2013)\citenamefont{Mohler, Prelovsek,
  and Woloshyn}}]{Mohler:2012na}
\bibinfo{author}{\bibfnamefont{D.}~\bibnamefont{Mohler}},
  \bibinfo{author}{\bibfnamefont{S.}~\bibnamefont{Prelovsek}},
  \bibnamefont{and} \bibinfo{author}{\bibfnamefont{R.}~\bibnamefont{Woloshyn}},
  \bibinfo{journal}{Phys. Rev. D} \textbf{\bibinfo{volume}{87}},
  \bibinfo{pages}{034501} (\bibinfo{year}{2013}).

\bibitem[{\citenamefont{Mahbub et~al.}(2013)\citenamefont{Mahbub, Kamleh,
  Leinweber, Moran, and Williams}}]{Mahbub:2012ri}
\bibinfo{author}{\bibfnamefont{M.~S.} \bibnamefont{Mahbub}},
  \bibinfo{author}{\bibfnamefont{W.}~\bibnamefont{Kamleh}},
  \bibinfo{author}{\bibfnamefont{D.~B.} \bibnamefont{Leinweber}},
  \bibinfo{author}{\bibfnamefont{P.~J.} \bibnamefont{Moran}}, \bibnamefont{and}
  \bibinfo{author}{\bibfnamefont{A.~G.} \bibnamefont{Williams}},
  \bibinfo{journal}{Phys. Rev. D} \textbf{\bibinfo{volume}{87}},
  \bibinfo{pages}{011501} (\bibinfo{year}{2013}).

\bibitem[{\citenamefont{Edwards et~al.}(2013)\citenamefont{Edwards, Mathur,
  Richards, and Wallace}}]{Edwards:2012fx}
\bibinfo{author}{\bibfnamefont{R.~G.} \bibnamefont{Edwards}},
  \bibinfo{author}{\bibfnamefont{N.}~\bibnamefont{Mathur}},
  \bibinfo{author}{\bibfnamefont{D.~G.} \bibnamefont{Richards}},
  \bibnamefont{and} \bibinfo{author}{\bibfnamefont{S.~J.}
  \bibnamefont{Wallace}}, \bibinfo{journal}{Phys. Rev. D}
  \textbf{\bibinfo{volume}{87}}, \bibinfo{pages}{054506}
  (\bibinfo{year}{2013}).

\bibitem[{\citenamefont{Moir et~al.}(2013)\citenamefont{Moir, Peardon, Ryan,
  Thomas, and Liu}}]{Moir:2013ub}
\bibinfo{author}{\bibfnamefont{G.}~\bibnamefont{Moir}},
  \bibinfo{author}{\bibfnamefont{M.}~\bibnamefont{Peardon}},
  \bibinfo{author}{\bibfnamefont{S.~M.} \bibnamefont{Ryan}},
  \bibinfo{author}{\bibfnamefont{C.~E.} \bibnamefont{Thomas}},
  \bibnamefont{and} \bibinfo{author}{\bibfnamefont{L.}~\bibnamefont{Liu}},
  \bibinfo{journal}{JHEP} \textbf{\bibinfo{volume}{1305}}, \bibinfo{pages}{021}
  (\bibinfo{year}{2013}).

\bibitem[{\citenamefont{Alexandrou et~al.}(2013)\citenamefont{Alexandrou,
  Korzec, Koutsou, and Leontiou}}]{Alexandrou:2013fsu}
\bibinfo{author}{\bibfnamefont{C.}~\bibnamefont{Alexandrou}},
  \bibinfo{author}{\bibfnamefont{T.}~\bibnamefont{Korzec}},
  \bibinfo{author}{\bibfnamefont{G.}~\bibnamefont{Koutsou}}, \bibnamefont{and}
  \bibinfo{author}{\bibfnamefont{T.}~\bibnamefont{Leontiou}}
  (\bibinfo{year}{2013}), \eprint{arXiv:1302.4410 [hep-lat]}.

\bibitem[{\citenamefont{Thomas et~al.}(2012)\citenamefont{Thomas, Edwards, and
  Dudek}}]{Thomas:2011rh}
\bibinfo{author}{\bibfnamefont{C.~E.} \bibnamefont{Thomas}},
  \bibinfo{author}{\bibfnamefont{R.~G.} \bibnamefont{Edwards}},
  \bibnamefont{and} \bibinfo{author}{\bibfnamefont{J.~J.} \bibnamefont{Dudek}},
  \bibinfo{journal}{Phys. Rev. D} \textbf{\bibinfo{volume}{85}},
  \bibinfo{pages}{014507} (\bibinfo{year}{2012}).

\bibitem[{\citenamefont{Morningstar and Peardon}(2004)}]{Morningstar:2003gk}
\bibinfo{author}{\bibfnamefont{C.}~\bibnamefont{Morningstar}} \bibnamefont{and}
  \bibinfo{author}{\bibfnamefont{M.~J.} \bibnamefont{Peardon}},
  \bibinfo{journal}{Phys. Rev. D} \textbf{\bibinfo{volume}{69}},
  \bibinfo{pages}{054501} (\bibinfo{year}{2004}).

\bibitem[{\citenamefont{Peardon et~al.}(2009)}]{distillation2009}
\bibinfo{author}{\bibfnamefont{M.}~\bibnamefont{Peardon}} \bibnamefont{et~al.},
  \bibinfo{journal}{Phys. Rev. D} \textbf{\bibinfo{volume}{80}},
  \bibinfo{pages}{054506} (\bibinfo{year}{2009}).

\bibitem[{\citenamefont{Moore and Fleming}(2006{\natexlab{a}})}]{Moore:2005dw}
\bibinfo{author}{\bibfnamefont{D.~C.} \bibnamefont{Moore}} \bibnamefont{and}
  \bibinfo{author}{\bibfnamefont{G.~T.} \bibnamefont{Fleming}},
  \bibinfo{journal}{Phys. Rev. D} \textbf{\bibinfo{volume}{73}},
  \bibinfo{pages}{014504} (\bibinfo{year}{2006}{\natexlab{a}}).

\bibitem[{\citenamefont{Moore and Fleming}(2006{\natexlab{b}})}]{Moore:2006ng}
\bibinfo{author}{\bibfnamefont{D.~C.} \bibnamefont{Moore}} \bibnamefont{and}
  \bibinfo{author}{\bibfnamefont{G.~T.} \bibnamefont{Fleming}},
  \bibinfo{journal}{Phys. Rev. D} \textbf{\bibinfo{volume}{74}},
  \bibinfo{pages}{054504} (\bibinfo{year}{2006}{\natexlab{b}}).

\bibitem[{\citenamefont{Lin et~al.}(2009)}]{Lin:2008pr}
\bibinfo{author}{\bibfnamefont{H.-W.} \bibnamefont{Lin}} \bibnamefont{et~al.}
  (\bibinfo{collaboration}{Hadron Spectrum}), \bibinfo{journal}{Phys. Rev. D}
  \textbf{\bibinfo{volume}{79}}, \bibinfo{pages}{034502}
  (\bibinfo{year}{2009}).

\bibitem[{\citenamefont{Clark et~al.}(2005)\citenamefont{Clark, Kennedy, and
  Sroczynski}}]{Clark:2004cp}
\bibinfo{author}{\bibfnamefont{M.~A.} \bibnamefont{Clark}},
  \bibinfo{author}{\bibfnamefont{A.~D.} \bibnamefont{Kennedy}},
  \bibnamefont{and}
  \bibinfo{author}{\bibfnamefont{Z.}~\bibnamefont{Sroczynski}},
  \bibinfo{journal}{Nucl.\ Phys.\ B (Proc.\ Suppl.)}
  \textbf{\bibinfo{volume}{140}}, \bibinfo{pages}{835} (\bibinfo{year}{2005}).

\bibitem[{\citenamefont{Foley et~al.}(2005)\citenamefont{Foley, Juge
  et~al.}}]{Foley:2005ac}
\bibinfo{author}{\bibfnamefont{J.}~\bibnamefont{Foley}},
  \bibinfo{author}{\bibfnamefont{K.~J.} \bibnamefont{Juge}},
  \bibnamefont{et~al.}, \bibinfo{journal}{Comput. Phys. Commun.}
  \textbf{\bibinfo{volume}{172}}, \bibinfo{pages}{145} (\bibinfo{year}{2005}).

\bibitem[{\citenamefont{Blossier et~al.}(2009)\citenamefont{Blossier,
  Della~Morte, von Hippel, Mendes, and Sommer}}]{Blossier:2009kd}
\bibinfo{author}{\bibfnamefont{B.}~\bibnamefont{Blossier}},
  \bibinfo{author}{\bibfnamefont{M.}~\bibnamefont{Della~Morte}},
  \bibinfo{author}{\bibfnamefont{G.}~\bibnamefont{von Hippel}},
  \bibinfo{author}{\bibfnamefont{T.}~\bibnamefont{Mendes}}, \bibnamefont{and}
  \bibinfo{author}{\bibfnamefont{R.}~\bibnamefont{Sommer}},
  \bibinfo{journal}{JHEP} \textbf{\bibinfo{volume}{0904}}, \bibinfo{pages}{094}
  (\bibinfo{year}{2009}).

\bibitem[{\citenamefont{Edwards and Joo}(2005)}]{Edwards:2004sx}
\bibinfo{author}{\bibfnamefont{R.~G.} \bibnamefont{Edwards}} \bibnamefont{and}
  \bibinfo{author}{\bibfnamefont{B.}~\bibnamefont{Joo}}
  (\bibinfo{collaboration}{SciDAC}), \bibinfo{journal}{Nucl. Phys. Proc.
  Suppl.} \textbf{\bibinfo{volume}{140}}, \bibinfo{pages}{832}
  (\bibinfo{year}{2005}).

\end{thebibliography}
\end{document}